\documentclass[preprint,12pt,compress]{elsarticle}

\usepackage{amsmath,amssymb,amsfonts}
\usepackage{graphics}
\usepackage{graphicx}
\usepackage{epstopdf, epsfig}
\usepackage{bm}
\usepackage{color}
\usepackage{hyperref}
\usepackage{comment}
\usepackage{mathrsfs}
\usepackage{libertine}
\usepackage{enumerate}
\usepackage[margin=2.4cm]{geometry}

\newcommand{\be}{\begin{equation}}
\newcommand{\ee}{\end{equation}}
\def\rr#1{(\ref{#1})}

\journal{Annals of Physics}

\begin{document}

\begin{frontmatter}

\title{$\mathscr{PT}$-symmetric hydrodynamics of odd viscous liquids and their oscillator counterparts}

\author[nw]{E. Kirkinis}
\author[uw]{A. Levchenko}

\address[nw]{Center for Computation and Theory of Soft Materials, Robert R. McCormick School of Engineering and Applied Science, Northwestern University, Evanston IL 60208 USA}
\address[uw]{Department of Physics, University of Wisconsin--Madison,
Madison, Wisconsin 53706, USA}

\date{\today}

\begin{abstract}
Odd viscosity, the nondissipative part of the viscous response of a time-reversal-broken fluid, is notoriously difficult to measure precisely because it does no work. Here we show that parity-time ($\mathscr{PT}$) symmetry, familiar from non-Hermitian optics, converts this elusiveness into a measurement principle. The odd Navier-Stokes equations, that include the nonlinear inertial terms, are $\mathscr{PT}$-symmetric, follow from a Lagrangian, and linearize to a Schr\"odinger equation in which the odd viscosity plays the role of Planck's constant; potential vorticity obeys a generalized Ertel conservation law. A probe trapped in an odd liquid realizes a pair of oscillators coupled by odd friction, and supplying balanced loss and gain drives a twofold $\mathscr{PT}$ transition whose exceptional point and Rabi sidebands locate the odd viscosity with square-root-enhanced sensitivity. Upon quantization the spectrum is of Fock-Darwin form, and the dissipative pair exhibits a Liouvillian exceptional point separating linear from exponential heating. These results furnish mechanical, stochastic, and spectroscopic protocols for measuring odd transport coefficients in classical and quantum fluids.
\end{abstract}

\begin{keyword}
Odd viscosity \sep $\mathscr{PT}$-symmetry \sep
Navier-Stokes equations \sep Ertel's theorem \sep
stochastic odd oscillator \sep quantum odd systems, Hamiltonian, compressible flow, incompressible flow
\end{keyword}

\end{frontmatter}

\tableofcontents

\section{Introduction}
Avron and coworkers pointed out that the hydrodynamic
equations are endowed with a second viscosity coefficient, termed odd or Hall viscosity, when time-reversal symmetry is broken either spontaneously or due to an external magnetic field or rotation \cite{Avron1995,Avron1998}. The Cauchy stress
tensor, apart from its usual rate-of-strain part, acquires additional terms, of non-dissipative nature.
There is a number of striking physical
manifestations associated with the presence of odd viscosity: a rotating disk experiences a normal compressive
or tensile stress \citep{Avron1998} in addition to the
shear stresses caused by the standard shear viscosity; a swimmer experiences a torque proportional to the rate-of-change of
its area \citep{Lapa2014}; an expanding bubble will promote an azimuthal flow on its surrounding liquid \citep{Ganeshan2017},
in addition to the radial flow existing in the presence of shear viscosity only;
chiral matter interfacial deformations level-off in a manner akin to surface tension \citep{Soni2019};
viscous unstable thin liquid films can become stabilized and a thermocapillary droplet can be set into motion \cite{kirkinis2019b,Aggarwal2023}. 
Odd viscosity was recently observed to persist and was experimentally measured in a colloidal liquid
consisting of spinning magnets \citep{Soni2019}; see \cite{Banerjee2017,Fruchart2023} for reviews in the context or chiral active matter.
It is worth emphasizing that odd viscosity was born quantum: it was first derived for quantum Hall fluids, where it is quantized \cite{Avron1995,Read2009}, a thread we take up again in section \ref{sec: quantum}.

Bateman \cite{Bateman1931} showed that a dissipative system could be derived from a Lagrangian density; see Morse and Feshbach \cite[Vol I, p.298 and onwards]{Morse1953} for an extended discussion. These ideas were employed later by Bender \cite{Bender2019} in both quantum mechanical and classical systems. In particular, it was shown that $\mathscr{PT}$-symmetric systems can exist as intermediaries between open and closed systems, and a twofold transition was established theoretically and experimentally in loss-gain resonators \cite{Bender2013,Peng2014}. The presence of 
$\mathscr{PT}$ symmetry, in general, improves the experimental sensitivity \cite{Chen2017} and has led to unconventional predictions in the area of optics, such as cloaking and 
loss-induced transparency, cf. \cite{Guo2009,Lin2011} and references therein.

Although $\mathscr{PT}$ symmetry has been established in many physical systems, experimentally realizing a number of practical consequences, whether $\mathscr{PT}$ symmetry is present in a fluid system and whether its consequences can be ascertained in an experiment, still remains an elusive accomplishment.  
In this paper, we provide this missing link 
by establishing the presence of the (exchange) $\mathscr{PT}$ symmetry according to Bender, in classical and quantum liquids and provide extensions that range from quantum Hall fluids and graphene electron hydrodynamics to chiral superfluids. This unified framework leads to mechanical, stochastic, and spectroscopic protocols for measuring odd transport coefficients in classical and quantum fluids.
 
\begin{figure}[b!]
\begin{center}
\includegraphics[height=2.7in,width=5.4in]{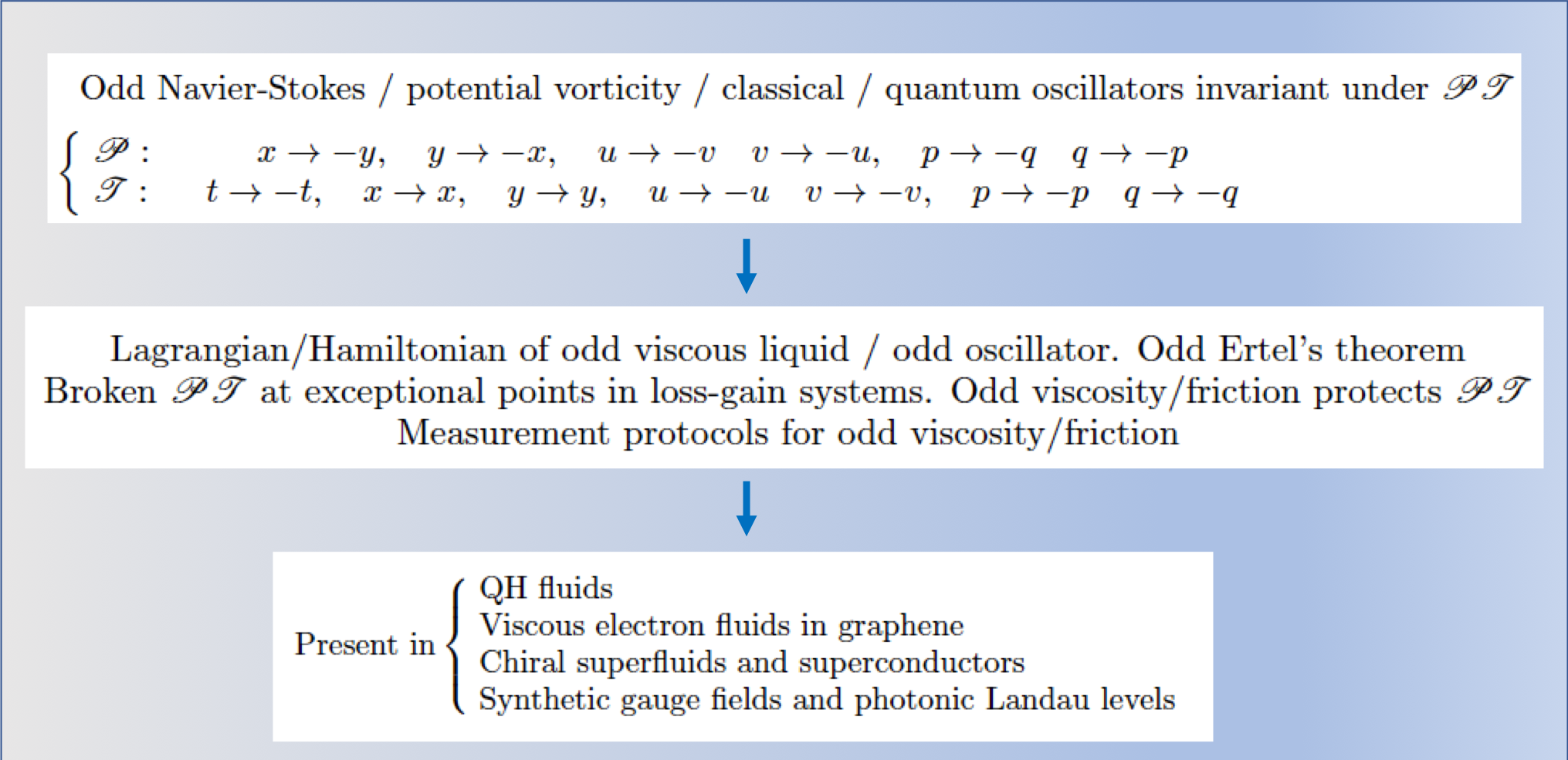}
\end{center}
\vspace{-5pt}
\caption{Summary of main concepts introduced in this paper
\label{message} }
\end{figure}

This article is organized as follows. In section \ref{sec: oddNS} we establish the invariance of the odd Navier-Stokes equations (which include the nonlinear inertial terms) under the combined action of $ \mathscr{PT}$ symmetry of a two-dimensional compressible or a three-dimensional incompressible liquid, whereby the $x$ and $y$ coordinates and momenta are interchanged by parity \cite{Bender2019}. A corollary of this formulation is that the maximal stresses in a liquid are also invariant, which could lead to observable consequences, cf. section \ref{sec: maxstress}. In section \ref{sec: Lagrangian} we show that the odd viscous terms can be derived from a Lagrangian density, exhibit the corresponding conserved Hamiltonian and Noether currents, and point out that in complex form the linearized bulk equations are of Schr\"odinger type by implementing the correspondence $\eta_o \leftrightarrow \hbar/2m$, where $\eta_o$ is an odd viscosity coefficient. In section \ref{sec: Ertel} we show how potential vorticity is conserved in a three-dimensional odd viscous liquid, thus giving rise to an extension of Ertel's theorem. In section \ref{sec: hydroPT} we demonstrate a genuinely hydrodynamic $\mathscr{PT}$ transition: transverse waves of a three-dimensional odd viscous liquid subject to gain and loss break $\mathscr{PT}$ symmetry below a critical wavenumber $k_\star=\sqrt{\gamma/\nu_4}$, where $\gamma$ and $\nu_4$ are the loss-gain and odd viscosity coefficients, respectively. Section \ref{sec: oscillators} treats what we call odd oscillators, that is, oscillators coupled through an odd friction coefficient $\beta$. Whereas odd forces based on displacement have been considered in active matter \cite{Caprini2025}, here the transverse force depends on velocities and can be expressed with respect to a Bateman-type Hamiltonian \cite{Bateman1931,Dekker1981}. We derive Lagrangians and Hamiltonians for odd oscillators, valid also in the presence of loss and gain, obtain exact laws for the Rabi frequencies, show that a probe trapped in an odd viscous liquid realizes this system with $\beta\propto\eta_o$ (section \ref{sec: probe}), discuss stochastic (section \ref{sec: stochastic}) and nonlinear (section \ref{sec: nonlinear}) extensions, and formulate the resulting measurement protocols for the odd friction and hence for odd viscosity. Section \ref{sec: quantum} is devoted to quantum odd systems: the exact Fock-Darwin spectrum of the quantized odd oscillator, the quantum master equation of the loss-gain pair and its Liouvillian exceptional point, and the connections to quantum Hall fluids, graphene electron hydrodynamics, chiral superfluids and synthetic gauge-field platforms. We conclude in section \ref{sec: conclusions}.

\section{\label{sec: oddNS}$ \mathscr{PT}$ symmetry of odd Navier-Stokes equations and conserved quantities}
For a viscous liquid in two dimensions endowed with odd viscosity, the stress tensor becomes
\cite{Avron1998,Lapa2014}
\be  \label{stress}
\sigma_{ik} = - \eta_o (\delta_{i1}\delta_{k1} - \delta_{i2}\delta_{k2} ) 
\left( \frac{\partial u_1}{\partial x_2} +\frac{\partial u_2}{\partial x_1}  \right)
+ \eta_o \left( \delta_{i1}\delta_{k2} + \delta_{i2} \delta_{k1}    \right) 
\left( \frac{\partial u_1}{\partial x_1} - \frac{\partial u_2}{\partial x_2}\right), 
\ee 
where $ i = 1,2$ and $k = 1,2$, and $\eta_o$
is the odd viscosity coefficient.
Eq. \rr{stress} acquires the simple matrix form
\be
\sigma = \eta_o\left( \begin{array}{cc}
 -(\partial_xv + \partial_y u) & \partial_x u  - \partial_yv \\
 \partial_x u  - \partial_yv & \partial_xv + \partial_y u
 \end{array} \right).
\ee

The Navier-Stokes equations in a frame rotating with constant angular velocity $\Omega$ about the $z$-axis (so that $-2\rho\Omega v$ and $+2\rho\Omega u$ are the components of the Coriolis force; the centrifugal contribution is absorbed into the pressure) obtain the form 
\begin{eqnarray}  \label{onsx}
\rho\left[ u_t + uu_x +vu_y\right] -2\rho\Omega v& = & -p_x  - \eta_o\nabla^2 v , \\
\rho\left[ v_t + uv_x +vv_y\right] +2\rho\Omega u& = & -p_y + \eta_o\nabla^2 u  \label{onsy}
\end{eqnarray}
and the continuity equation reads
\be \label{cont1}
\partial_t \rho' + \rho\,\textrm{div} \mathbf{v} =0 \quad \textrm{for} \quad \rho'\ll\rho,
\ee
where $\rho'$ is the variable part of the density, $\rho$ a constant background level, $p= \rho c^2$ and $c$ is the speed of sound.

Eqs. \rr{onsx}-\rr{cont1}
are invariant under the combined action of parity and time-reversal (i.e. they are $ \mathscr{PT}$ - symmetric):
Under parity (exchange) the coordinates are interchanged and reflected
\be \label{P1}
\mathscr{P}: \quad x \rightarrow - y, \quad y \rightarrow -x, \quad u \rightarrow -v \quad v \rightarrow -u.
\ee
Under time-reversal the momenta are reversed
\be \label{T1}
\mathscr{T}: \quad t \rightarrow -t, \quad x \rightarrow x, \quad y \rightarrow y, \quad u \rightarrow -u \quad v \rightarrow -v.
\ee
Equations \rr{onsx}-\rr{cont1} are not invariant under $\mathscr{P}$ or $\mathscr{T}$ separately \footnote{The negative sign in Eq. \rr{P1} is included only for compatibility with the standard parity operation, see also \cite[Ref.25]{Bender2013}}. Invariance under the combined operation, including the Coriolis terms and the continuity equation, follows by direct substitution of $u'(x,y,t) = v(-y,-x,-t)$, $v'(x,y,t) = u(-y,-x,-t)$, $p'(x,y,t)=p(-y,-x,-t)$.
The sign of $\Omega$ remains unchanged\footnote{For instance, a vector making an angle $\theta(t)$ with the $x$-axis and rotating counterclockwise in the $x$-$y$ plane, is mapped by parity \rr{P1} to a vector, rotating clockwise, making an angle $3\pi/2 -\theta$ with the same axis. Thus, under the combined
$ \mathscr{PT}$ symmetry \rr{P1}, \rr{T1}, $d\theta/dt$ does not change sign.}.

It is instructive to combine the two velocity components into the complex field $\psi = u + iv$. Equations \rr{onsx}-\rr{onsy} then read
\be \label{schrod}
i\left(\partial_t + \mathbf{v}\cdot\nabla + 2i\Omega\right)\psi = -\nu_o \nabla^2 \psi + \frac{i}{\rho}(\partial_x + i \partial_y) p,
\ee
with $\nu_o = \eta_o/\rho$ the kinematic odd viscosity. Linearized about the state of rest ($\Omega=0$, uniform pressure), Eq. \rr{schrod} is the free Schr\"odinger equation for the complex velocity, with the correspondence
\be \label{hbar}
\nu_o \longleftrightarrow \frac{\hbar}{2m}.
\ee
The odd ``viscous'' term is therefore dispersive rather than dissipative, it does no work (see also section \ref{sec: Lagrangian}), and the correspondence \rr{hbar} anticipates the quantum considerations of section \ref{sec: quantum}.

In three dimensions the odd Navier-Stokes equations read
\be \label{NS2}
\frac{d u}{d t} = - \frac{1}{\rho}\frac{\partial \tilde{p}}{\partial x}  -\mathcal{S} v, \quad \frac{d v}{d t} = -\frac{1}{\rho}\frac{\partial \tilde{p}}{\partial y} + \mathcal{S} u, \quad \frac{d w}{d t} = -\frac{1}{\rho}\frac{\partial \tilde{p}}{\partial z},
\ee
where the operator
\be \label{S}
\mathcal{S} = (\nu_o-\nu_4)\nabla^2_2 + \nu_4\partial_z^2, 
\ee
is built from the two odd kinematic viscosity coefficients $\nu_o$ and $\nu_4$, cf. \cite{Kirkinis2023taylor,kirkinis2024}, and \cite{Landau1981}; $\nabla^2_2$ is the two-dimensional (horizontal) Laplacian, $d/dt$ denotes the convective derivative and $\tilde{p}$ is a modified pressure, see the discussion in \ref{sec: 3D} for the derivation of Eq. \rr{NS2} and its notation. It is clear that they are invariant with respect to the
$ \mathscr{PT}$ symmetry \rr{P1}, \rr{T1} in three dimensions that includes the parity $z\rightarrow -z$ and time-reversal. In addition, in three dimensions a more symmetric counterpart of the $ \mathscr{PT}$ symmetry \rr{P1}, \rr{T1} can be introduced that takes the form 
\be \label{P3}
\mathscr{P}: \quad x \rightarrow - y, \quad y \rightarrow -z,\quad z\rightarrow -x, \quad u \rightarrow -v \quad v \rightarrow -w, \quad w \rightarrow -u
\ee
and
\be \label{T3}
\mathscr{T}: \quad t \rightarrow -t, \quad x \rightarrow x, \quad y \rightarrow y, \quad z \rightarrow z, \quad u \rightarrow -u \quad v \rightarrow -v\quad w \rightarrow -w,
\ee
giving rise to \rr{NS2} by a suitable relabeling of the coordinates and fields.

\subsection{\label{sec: maxstress}Maximal stresses}
Let $\hat{\mathbf{n}} = (n_1,n_2)$, and $\hat{\mathbf{t}}=(-n_2,n_1)$ be the normal and tangent vectors at a surface in the two-dimensional incompressible liquid. The normal and shear stresses can be
written as
\be
\hat{\mathbf{n}} \sigma \hat{\mathbf{n}}  = \sigma_1 n_1^2 + \sigma_2 n_2^2, \quad
|\hat{\mathbf{t}} \sigma \hat{\mathbf{n}} | = | (\sigma_1 - \sigma_2)n_1n_2|
\ee
where
\be \label{sigma12}
\sigma_{1,2} = \pm \eta_o\sqrt{(u_x-v_y)^2 + (u_y+v_x)^2}
\ee
are the principal stresses.
Following \cite[p.103]{Chadwick1976}, it is straightforward to show that the extremal values of the normal stress
are the principal stresses \rr{sigma12} and that $\hat{\mathbf{n}}  \sigma \hat{\mathbf{n}}  = \sigma_i$ when $\hat{\mathbf{n}} $ is aligned with
the corresponding principal axis of stress.

The maximal value of the shear stress is
\be \label{shearmax}
|\hat{\mathbf{t}} \sigma \hat{\mathbf{n}} | = \frac{1}{2}|\sigma_1 - \sigma_2| = \eta_o \sqrt{(u_x-v_y)^2 + (u_y+v_x)^2}
\ee
obtained when the normal directions are
inclined at either $\pi/4$ or $3\pi/4$ with respect to the principal axes. In this state,
the normal stress
$
\hat{\mathbf{n}} \sigma \hat{\mathbf{n}}  = \frac{1}{2}(\sigma_1 + \sigma_2)
$
vanishes.

Equation \rr{shearmax} is invariant under the combined $ \mathscr{PT}$ symmetry \rr{P1} and \rr{T1}. This implies that a measurement of the stress at point $(x,y)$ is the same as at the point $(y,x)$ with a suitable choice of velocities, which could be employed to enable the measurement of the odd viscosity coefficient; a transport analogue of this statement for electron fluids is formulated in section \ref{sec: graphene}.

It is also easy to show that when shear and bulk viscosities
$\eta$ and $\zeta$ are included along with the odd constitutive law \rr{stress}, the principal stresses acquire the simple form
\be \label{sigma12shear}
\sigma_{1,2} =\zeta(u_x+v_y) \pm (\eta + \eta_o)^{1/2}\sqrt{(u_x-v_y)^2 + (u_y+v_x)^2}. 
\ee
The combined $ \mathscr{PT}$ symmetry in this case exchanges the role of the principal stresses. The above conlcusions remain unchanged with the exception of the non-vanishing normal stress 
$
\hat{\mathbf{n}} \sigma \hat{\mathbf{n}} 
$.

\subsection{\label{sec: Lagrangian}Lagrangian, Hamiltonian \& Noether currents}
The Navier-Stokes equations \rr{onsx} and \rr{onsy} can be derived from a Lagrangian density $\mathcal{L}$\footnote{\label{nonuniqueness} The term generating the nonlinear contributions is unique up to a total
divergence; see \ref{sec: nullapp}.
}
\be \label{Ldensity}
\mathcal{L} = \frac{1}{2}\left(u v_t - vu_t\right) + \frac{\eta_o}{2} \left(u_x^2 + u_y^2 + v_x^2 + v_y^2\right)
+\frac{1}{2}\left( u^2 v_x - v^2 u_y\right) + \Omega\left(u^2 + v^2\right),
\ee
where we set $\rho=1$; the last term generates the Coriolis force.
The Euler-Lagrange equation for $u$
\be
\frac{\partial}{\partial t} \left(\frac{\partial \mathcal{L}}{\partial u_t}\right)
 + \frac{\partial}{\partial x} \left(\frac{\partial \mathcal{L}}{\partial u_x}\right)
+ \frac{\partial}{\partial y} \left(\frac{\partial \mathcal{L}}{\partial u_y}\right)
- \left(\frac{\partial \mathcal{L}}{\partial u}\right) = 0
\ee
gives \rr{onsy} and the Euler-Lagrange equation for $v$ gives \rr{onsx} (for incompressible flow, the pressure is introduced
as a Lagrange multiplier that incorporates the incompressibility condition, cf. \cite{Serrin1959}).

If by $p$ and $q$ we denote the generalized momenta conjugate to $u$ and $v$ respectively, then since $p = \frac{\partial \mathcal{L}}{\partial u_t} = -\frac{1}{2} v$ and
$q = \frac{\partial \mathcal{L}}{\partial v_t} = \frac{1}{2} u$, the corresponding Hamiltonian density reads
\be \label{H0}
\mathcal{H} = pu_t + q v_t -\mathcal{L} = -\frac{\eta_o}{2} \left( |\nabla u|^2 +  |\nabla v|^2 \right) -  \frac{1}{2}\left( u^2 v_x - v^2 u_y\right) - \Omega\left(u^2+v^2\right),
\ee
where $|\nabla u|^2 = u_x^2 + u_y^2$ etc. (Note the overall minus sign relative to the non-kinetic part of $\mathcal{L}$: since the kinetic term of \rr{Ldensity} is linear in the time derivatives, it cancels in the Legendre transform, which then returns minus the remaining terms. The overall sign of $\mathcal{L}$, and hence of $\mathcal{H}$, carries no physical significance.) The Hamilton equations can be written as in
\cite[Eq. (3.3.24)]{Morse1953}; only two
Hamilton equations are generated and these
are just \rr{onsx} and \rr{onsy}.
The important feature is that $H = \int \mathcal{H}\, dxdy$ remains constant as the system evolves in time: the odd viscous term, despite containing gradients of the velocity, does no work.
The Hamiltonian \rr{H0} is invariant with respect to the combined $\mathscr{P}$ and
$\mathscr{T}$ transformations in \rr{P1} and \rr{T1}, but is not invariant with respect to
$\mathscr{P}$ and
$\mathscr{T}$ separately.

Since $\mathcal{L}$ does not depend explicitly on $x,y$, Noether's theorem supplies, in addition to $H$, the conserved momenta $P_i = \int \mathcal{P}_i \,dxdy$ with densities
\be \label{noether}
\mathcal{P}_i = \frac{\partial\mathcal{L}}{\partial u_t}\,\partial_i u + \frac{\partial\mathcal{L}}{\partial v_t}\,\partial_i v = \frac{1}{2}\left( u\, \partial_i v - v\,\partial_i u\right) = \frac{1}{2}\,\textrm{Im}\left( \bar{\psi}\, \partial_i \psi \right), \qquad i = x,y,
\ee
which is precisely the Schr\"odinger current of the complex velocity $\psi = u+iv$ introduced in \rr{schrod}, reinforcing the correspondence \rr{hbar}. The associated fluxes follow from the canonical energy-momentum tensor $T_{ij} = \partial\mathcal{L}/\partial(\partial_j \phi_a)\, \partial_i \phi_a - \delta_{ij}\mathcal{L}$, $\phi_a \in \{u,v\}$. 

The density \rr{Ldensity} is not unique, but the ambiguity can be stated
precisely. Within the class of terms cubic in the velocities and of first
order in their gradients, the generating term
$\frac{1}{2}(u^2v_x - v^2u_y)$ is determined exactly up to a null
Lagrangian, a total divergence $D_x G(u,v) + D_y H(u,v)$ with $G, H$
arbitrary homogeneous cubics, and no further freedom arises
(\ref{sec: nullapp}). All admissible representatives therefore
yield identical field equations, identical canonical momenta, and the same
integrated charges $H$ and $P_i$ under decaying or periodic boundary
conditions; only the local densities and boundary fluxes are
representative-dependent, a distinction that acquires physical content on
bounded domains, where odd liquids support edge dynamics \cite{Abanov2018}.
\ref{sec: nullapp} also records why a variational principle exists
at all, given the classical obstructions to Eulerian variational
formulations of the Navier-Stokes equations
\cite{Millikan1929,Finlayson1972}: in the symplectic pairing of $u$ with
$v$ selected by the kinetic term of \rr{Ldensity}, the odd viscous operator
is self-adjoint in the Helmholtz sense, whereas the shear viscous operator
is not, which is why even viscosity requires the doubling of degrees of
freedom introduced by Bateman \cite{Bateman1931}, while odd viscosity does
not.

In three dimensions the odd viscous term in \rr{Ldensity} is replaced by
\be \label{Ldensity3}
\mathcal{L}_{odd} =(\eta_o - \eta_4)\frac{1}{2} \left(u_x^2 + u_y^2 + v_x^2 + v_y^2\right)
+\frac{\eta_4}{2}\left( u^2_z  +  v^2_z \right),
\ee
but, in general, a full Lagrangian like \rr{Ldensity} is not readily available. 

\subsection{\label{sec: Ertel}Odd Ertel's theorem; potential vorticity - a conserved quantity}
Chiefly met in physical oceanography, potential vorticity is a vorticity-related scalar that takes into account liquid inhomogeneities \cite{Haynes1987,Haynes1990}.
Various geophysical applications of potential vorticity are discussed by Pedlosky \cite{Pedlosky1987},
while Landau \& Lifshitz  \cite[Problem \S 8]{Landau1987} discuss the related non-isentropic evolution of entropy in an ideal fluid.
Conservation of potential vorticity can be traced to the particle-relabeling symmetry of hydrodynamic action principles \cite{Salmon1988,Muller1995}; the derivation below proceeds instead directly from the equations of motion following the discussion of \cite{Muller1995} and \cite[p. 370]{Acheson1990}. 

Consider a three-dimensional odd viscous liquid with odd viscosity coefficients $\nu_o$ and $\nu_4$ \cite{Kirkinis2023taylor,kirkinis2024} and, in addition, shear viscosity $\nu$, so that the right-hand sides of \rr{NS2} acquire the familiar terms $\nu\nabla^2(u,v,w)$. Let $\lambda(\mathbf{x},t)$ be a scalar field carried by the flow---for instance the density in a Boussinesq liquid, the entropy, or the concentration of a passive tracer. The statements \rr{eqpv0}-\rr{cons1} below hold for arbitrary $\lambda$. 

The vorticity equation
\be \label{vort0}
\frac{D }{Dt}   \textrm{curl}\mathbf{v} = ( \textrm{curl}\mathbf{v} \cdot \nabla) \mathbf{v} + \nu \nabla^2  \textrm{curl}\mathbf{v} - \partial_z \mathcal{S} (\hat{\mathbf{x}}u + \hat{
\mathbf{y}} v),
\ee
is obtained by taking the curl of \rr{NS2} augmented by shear viscosity (the odd force $(-\mathcal{S}v, \mathcal{S}u,0)$ has curl $-\partial_z \mathcal{S}(\hat{\mathbf{x}}u + \hat{
\mathbf{y}} v)$ for incompressible flow, see Eq. \rr{S}). Multiply by
$\nabla \lambda$ to obtain
\be \label{eqpv0}
\frac{D }{Dt} (\bm{\omega} \cdot \nabla \lambda ) =   (\bm{\omega} \cdot \nabla) \frac{D\lambda }{Dt}
 + ( \nu \nabla^2 \bm{\omega} -  \partial_z \mathcal{S} (\hat{\mathbf{x}}u + \hat{
\mathbf{y}} v)) \cdot \nabla \lambda, 
\ee
where $\bm{\omega} = \textrm{curl}\mathbf{v}$.

Let
\be \label{PV}
q = \bm{\omega} \cdot \nabla \lambda
\ee
be the potential vorticity. Then, \rr{eqpv0} can be written as a conservation law
\be \label{cons1}
\partial_t q + \nabla \cdot \mathbf{F} =0, \quad \textrm{where} \quad \mathbf{F} =  q \mathbf{v} -  \lambda  \left[ \nu \nabla^2  \textrm{curl}\mathbf{v} - \partial_z \mathcal{S} (\hat{\mathbf{x}}u + \hat{
\mathbf{y}} v)
\right]-\bm{\omega} \frac{D \lambda}{Dt}.
\ee
From \rr{Fcdot} of \ref{sec: ErtelApp} we obtain
\be \label{cfluidvol}
\partial_t \lambda + \frac{\mathbf{F}}{q} \cdot \nabla \lambda =0,
\ee
which recovers Eq. (26) of \cite{Muller1995}. This relation
implies that $\lambda$ is a constant for a fluid volume moving with velocity $\mathbf{F}/q$.  
We emphasize the physical content of \rr{cons1}: potential vorticity is not materially conserved in an odd viscous liquid ($\mathbf{F} \neq q\mathbf{v}$); instead, it obeys a local conservation law whose flux contains the odd viscous contribution $\lambda\, \partial_z \mathcal{S}(\hat{\mathbf{x}}u + \hat{
\mathbf{y}} v)$. In flows where the flow is confined---for example the Taylor-column configurations of \cite{Kirkinis2023taylor}---the integral $\int q \,dV$ over a material volume is conserved and constrains the dynamics in the same manner as in rotating geophysical flows. The above discussion can be repeated for the potential vorticity in terms of the more general absolute vorticity observable $\bm{\omega} + 2\bm{\Omega}$, where $\bm{\Omega}$ is the angular velocity of the rotating coordinate system. 

In the absence of shear viscosity $\nu \equiv 0$, relations \rr{cons1} and \rr{cfluidvol} are invariant under the combined $\mathscr{PT}$ symmetry. Relations \rr{cons1} and \rr{cfluidvol} provide an additional body of tests to existing criteria \cite{kirkinis2024, Kirkinis2023halos,Kirkinis2025wave} for the determination of the largely unknown viscosity coefficients. 

\subsection{\label{sec: hydroPT}A hydrodynamic $\mathscr{PT}$ transition: odd transverse waves with gain and loss}
\begin{figure}
\begin{center}
\includegraphics[width=3.2in]{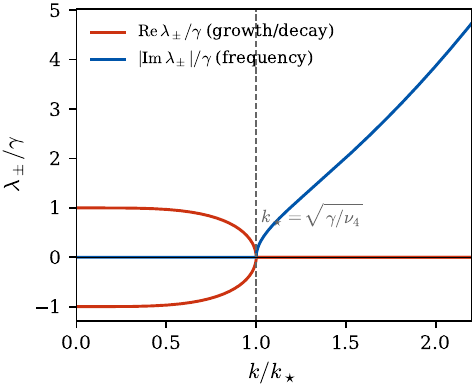}
\end{center}
\vspace{-5pt}
\caption{$\mathscr{PT}$ transition of transverse waves in a three-dimensional odd viscous liquid with balanced gain and loss of the two polarizations, Eq. \rr{oddwavegl}. Long-wavelength modes ($k<k_\star$, cf. Eq. \rr{kstar}) form growing/decaying pairs (broken $\mathscr{PT}$); short-wavelength modes oscillate at frequency $\sqrt{\nu_4^2k^4-\gamma^2}$ (unbroken $\mathscr{PT}$). The exceptional point of each mode is located by the odd viscosity $\nu_4$.
\label{fig: hydroPT} }
\end{figure}

In section \ref{sec: oscillators} we introduce and discuss extensively 
the notion of odd oscillators coupled by an odd friction coefficient $\beta$ and carrying a loss and gain coefficient $\gamma$. 
The oscillator systems of section \ref{sec: oscillators} are not merely analogues: they arise as single-mode reductions of the odd hydrodynamic equations. To see this, linearize \rr{NS2} about the state of rest and consider transverse plane waves $\mathbf{v} = (u(t), v(t), 0)\, e^{ikz}$ propagating along $z$; such flows are exactly incompressible and pressure-free. Since $\mathcal{S}e^{ikz} = -\nu_4 k^2 e^{ikz}$,
\be \label{oddwave}
\dot{u} = \nu_4 k^2\, v, \qquad \dot{v} = -\nu_4 k^2\, u,
\ee
i.e. the two transverse polarizations rotate into one another at the rate $\nu_4 k^2$: these are the circularly polarized transverse waves of the odd liquid \cite{kirkinis2024}. Now endow the two polarizations with gain and loss at rate $\gamma$ (in a chiral active liquid, energy injection by the spinners can be biased along one direction, e.g. by anisotropic substrate friction, while the other polarization is damped):
\be \label{oddwavegl}
\dot{u} = \gamma u + \nu_4 k^2 v, \qquad \dot{v} = -\gamma v - \nu_4 k^2 u .
\ee
The system \rr{oddwavegl} is $\mathscr{PT}$-symmetric under \rr{P1}-\rr{T1} and its eigenvalues
\be \label{hydroev}
\lambda_\pm = \pm\sqrt{\gamma^2 - \nu_4^2 k^4}
\ee
exhibit a $\mathscr{PT}$ transition at the critical wavenumber
\be \label{kstar}
k_\star = \sqrt{{\gamma}/{\nu_4}} :
\ee
short waves, $k>k_\star$, oscillate with frequency $\sqrt{\nu_4^2k^4 - \gamma^2}$ (unbroken phase; the odd coupling shuttles energy between the amplified and damped polarizations fast enough for neither to grow), while long waves, $k<k_\star$, break into a growing-decaying pair (broken phase); see Fig. \ref{fig: hydroPT}. Every mode $k$ sits at a distance from its exceptional point set by the odd viscosity, so a measurement of $k_\star$ measures $\nu_4$. Equations \rr{oddwavegl} are precisely of the form to which the loss-gain odd oscillator of section \ref{sec: frictionlossgain} reduces in the rotating-wave approximation (section \ref{sec: lindblad}), with the identification $\beta/2 \leftrightarrow \nu_4 k^2$. Setting $\nu_4k_*^2 =\beta$, Eq. \rr{kstar} leads the exceptional point to occur at $\gamma  =\beta$, which coincides with the corresponding oscillator interpretation point (i) in section \ref{sec: frictionlossgain}. 

In the absence of loss and gain, we can also interpret \rr{hydroev} as a dispersion relation which recovers its counterpart derived from an inertial-like waves formulation of odd viscous liquids, cf. the discussion in \cite[\S 8.3]{Kirkinis2023taylor}, setting $k=k_z$ and $\theta=0$ in \cite[Eq. (8.7)]{Kirkinis2023taylor}. 

\section{\label{sec: oscillators}$ \mathscr{PT}$ symmetry of odd oscillators}

In this section we consider systems of oscillators endowed with odd friction $\beta$, that is, oscillators
of the form
\be \label{system0}
\ddot{\mathbf{x}} +  \left( \begin{array}{cc}
 0& -\beta \\
 \beta & 0
 \end{array} \right) \dot{\mathbf{x}} + \omega^2 \mathbf{x} = \mathbf{0}
\ee
where $\mathbf{x} = (x,y)^T$. They are $\mathscr{P\!T}$-symmetric according to Bender \cite{Bender2013,Peng2014,Bender2019} and the equations of motion can be derived from a Lagrangian. We show that a loss-gain resonator with odd friction exhibits a twofold transition, obtain exact laws for its Rabi frequencies, and construct its Hamiltonian, which degenerates precisely at the exceptional point. The definition of what constitutes an ``odd oscillator'' differs from analogous interpretations in the context of odd active solids \cite{Caprini2025}. In that reference the odd terms are proportional to displacements; here they are proportional to velocities. Physical realizations of \rr{system0} include a charged particle in a magnetic field with $\beta$ the cyclotron frequency (see below), gyroscopic pendula \cite{Nash2015}, and a probe trapped in an odd viscous liquid (section \ref{sec: probe}).

\subsection{\label{sec: pureodd}Pure odd oscillator}
Coupling between the two fields $x$ and $y$ is provided by odd friction $\beta$:
\be \label{system1}
 \ddot{x}  - \beta \dot{y} + \omega^2 x = 0, \quad
 \ddot{y} + \beta \dot{x} + \omega^2 y = 0.
\ee
The characteristic polynomial is
$
P_1(\lambda) = \lambda^{4}+\left(\beta^{2}+2 \omega^{2}\right) \lambda^{2}+\omega^{4}.
$
The eigenvalues of this oscillator are purely imaginary
\be \label{evs1}
\lambda_k=\pm\frac{i}{2} \sqrt{2 \beta^{2}+4 \omega^{2}\pm 2 \sqrt{\beta^{4}+4 \beta^{2} \omega^{2}}} ,
\quad k = 1,2,3,4,
\ee
or, more transparently, $\lambda = \pm i f_\pm$ with the two positive frequencies
\be \label{fpm}
f_\pm = \sqrt{\omega^2 + \frac{\beta^2}{4}} \pm \frac{\beta}{2}, \qquad f_+ - f_- = \beta, \qquad f_+ f_- = \omega^2.
\ee
Equations \rr{system1} coincide with those of a particle of unit mass and unit charge in a harmonic trap $\omega$ subject to a uniform magnetic field of cyclotron frequency $\beta$: odd friction is a Lorentz-like, work-free force, and $f_\pm$ are the classical Fock-Darwin frequencies. This observation determines the Hamiltonian below and is the seed of the quantum treatment in section \ref{sec: fockdarwin}.

Following Morse and Feshbach \cite[Vol I, p.298]{Morse1953}, or directly from the magnetic analogy in the symmetric gauge, we can write the Hamiltonian
\be \label{H1}
H_1= \frac{1}{2}\left(p^2 + q^2\right) + \frac{\beta}{2}(py - qx) + \frac{1}{2}\left( \omega^2 + \frac{\beta^2}{4}\right)(x^2 + y^2).
\ee
(Note the factors $\beta/2$ and $\beta^2/4$: the symmetric gauge carries half the cyclotron frequency. A Hamiltonian with cross term $\beta(py-qx)$ and potential $\frac12(\beta^2+\omega^2)(x^2+y^2)$ generates \rr{system1} with $\beta \rightarrow 2\beta$ instead.)
The Hamiltonian \rr{H1} is $\mathscr{PT}$-symmetric. Under parity, the two degrees of freedom are exchanged
\be \label{P11}
\mathscr{P}: \quad x \rightarrow - y, \quad y \rightarrow -x, \quad p \rightarrow -q \quad q \rightarrow -p
\ee
Under time-reversal the momenta are reversed
\be \label{T11}
\mathscr{T}: \quad x \rightarrow x, \quad y \rightarrow y, \quad p \rightarrow -p \quad q \rightarrow -q
\ee
Hamilton's equations are
\be \label{Ham1a}
\dot{x} = \frac{\partial H_1}{\partial p} = p + \frac{\beta}{2} y, \quad \dot{p} = -\frac{\partial H_1}{\partial x} = \frac{\beta}{2} q - \left(\omega^2 + \frac{\beta^2}{4}\right) x,
\ee
and
\be \label{Ham1b}
\dot{y} = \frac{\partial H_1}{\partial q} = q - \frac{\beta}{2} x,
 \quad
\dot{q} = -\frac{\partial H_1}{\partial y} = -\frac{\beta}{2} p - \left(\omega^2+\frac{\beta^2}{4}\right) y,
\ee
which reproduce \rr{system1}. Expressed in terms of velocities, $H_1$ becomes the (obvious) first integral
$H_1 = \frac{1}{2}[\dot{x}^2+ \dot{y}^2 + \omega^2(x^2 + y^2)] = E_1$: odd friction does no work, and the energy of the oscillator pair is conserved.
The Lagrangian $L = p\dot{x} + q\dot{y} - H_1$ reads
\be \label{L2}
L = \frac{1}{2}(\dot{x}^2 + \dot{y}^2) - \frac{\beta}{2} (\dot{x}y - x\dot{y}) - \frac{1}{2}\omega^2 (x^2 + y^2).
\ee
Note that terms coupling velocities to positions as in \rr{L2} were recently introduced in dispersive parity breaking of a dielectric in a one-dimensional setting, cf. \cite[\S 4.1 \& 4.2]{Srivastava2024}, and they are the characteristic ingredient of the Bateman Lagrangian \cite{Bateman1931,Dekker1981}.

Expressing Eq. \rr{system1} as a first order system of four equations $\dot{\mathbf{u}} = A\mathbf{u}$, we can write the solution explicitly with
respect to the exponential matrix, $ \mathbf{u} = e^{At} \mathbf{u}_0$, where $\mathbf{u}_0$ is the four-dimensional vector of initial conditions. Instead, it is more instructive to treat $\beta$ as a perturbation
parameter and expand the solution of \rr{system1} in a regular perturbation expansion (setting $\omega = 1$). Eliminating secular
terms we obtain the pair of complex amplitude equations
\be \label{amp1}
\dot{A} = \frac{\beta}{2} B + \frac{i\beta^2}{8}A + O(\beta^3), \quad \dot{B} = -\frac{\beta}{2} A + \frac{i\beta^2}{8}B +O(\beta^3),
\ee
where $A$ and $B$ are slowly-varying amplitudes appearing in the uniformly valid expansion
$x(t) = A(\beta t)e^{i\omega t} + c.c.$ and $y(t) = B(\beta t)e^{i\omega t} + c.c.$ (the common phase $i\beta^2/8$ is the second-order frequency shift contained in the exact counterpart $f_\pm = 1 \pm \beta/2 + \beta^2/8 + \ldots$ of \rr{fpm}). This implies that the
solution has the form
\be
x(t) = 2R \cos(t+\theta), \quad y(t) = 2\rho \cos(t +\phi)
\ee
for $A=Re^{i\theta}$ and $B = \rho e^{i\phi}$. The amplitude equations \rr{amp1} have the first integrals
\be \label{firstint}
R^2 + \rho^2 = C_1^2, \quad \rho R \sin (\phi - \theta) = C_2
\ee
determined from the real counterparts of \rr{amp1}
\be \label{amp2a}
\dot{R} = \frac{\beta}{2}\rho \cos(\phi - \theta), \quad \dot{\rho} = -\frac{\beta}{2}R \cos(\phi - \theta),
\ee
and
\be\label{amp2b}
\dot{\theta} = \frac{\beta}{2}\frac{\rho}{R} \sin(\phi - \theta) + \frac{\beta^2}{8}, \quad \dot{\phi} = \frac{\beta}{2}\frac{R}{\rho} \sin(\phi - \theta) + \frac{\beta^2}{8}.
\ee
The pair $(A,B)$ rotates rigidly at the Rabi rate $\beta/2$: energy is completely transferred from the $x$ to the $y$ oscillator and back with period $2\pi/\beta$, consistent with the exact splitting $f_+-f_-=\beta$ of \rr{fpm}.

\begin{figure}
\begin{center}
\includegraphics[height=1.3in,width=6in]{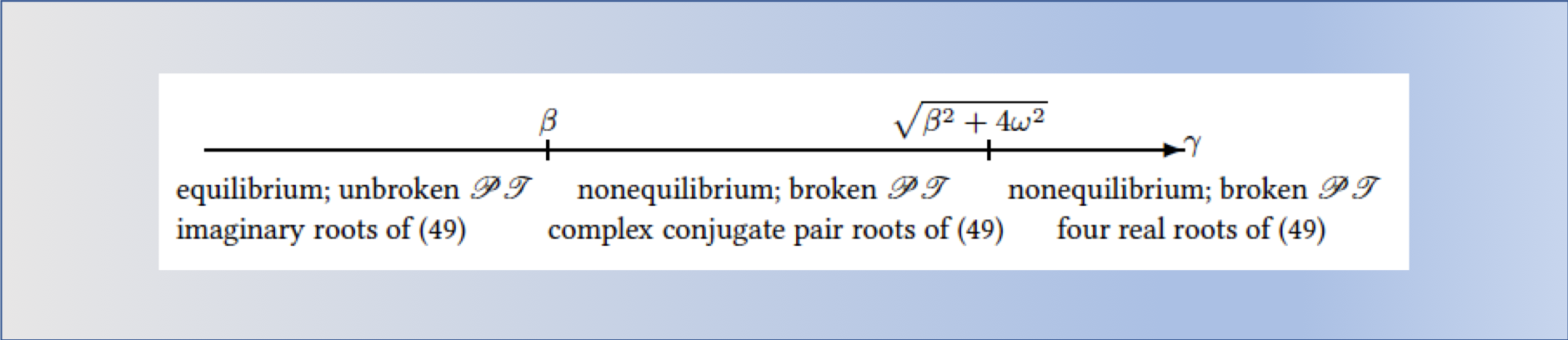}
\end{center}
\vspace{-1pt}
\caption{Regimes that determine the character of the eigenvalues \rr{evs2}.
In the left-most region the sign of $(\gamma^2 - \beta^2)(\gamma^2 - \beta^2 - 4\omega^2)$ is positive
and the eigenvalues \rr{evs2} are imaginary (unbroken $\mathscr{PT}$). In the middle region the sign of $(\gamma^2 - \beta^2)(\gamma^2 - \beta^2 - 4\omega^2)$ is negative and the eigenvalues become two complex conjugate pairs (broken $\mathscr{PT}$, oscillatory).
In the right-most region the sign of $(\gamma^2 - \beta^2)(\gamma^2 - \beta^2 - 4\omega^2)$ is positive and the eigenvalues are real (broken $\mathscr{PT}$, overdamped).
\label{fig1} }
\vspace{0pt}
\end{figure}

\subsection{\label{sec: frictionlossgain}Odd friction and loss-gain oscillators: a twofold transition}
We are interested here in the effect odd friction has on loss-gain systems.
Consider the coupled
system
\begin{align} \label{system2a}
 \ddot{x} + \gamma \dot{x} - \beta \dot{y} + \omega^2 x &= 0\\
 \ddot{y}  - \gamma\dot{y}+ \beta \dot{x} + \omega^2 y &= 0 \label{system2b}
\end{align}
with loss $\gamma>0$ acting on $x$ and equal gain acting on $y$. The characteristic polynomial is
\be \label{plambda}
P_2(\lambda) = \lambda^{4}+\left(\beta^{2}-\gamma^{2}+2 \omega^{2}\right) \lambda^{2}+\omega^{4}
\ee
with eigenvalues
\be \label{evs2}
\lambda_k=\pm\frac{1}{\sqrt{2}} \sqrt{\gamma^2 - \beta^{2} -2 \omega^{2}\pm \sqrt{(\gamma^2 - \beta^2)(\gamma^2 - \beta^{2} -4\omega^{2})}} ,
\ee
for $ k = 1,2,3,4$.

The central observation is the following. At $\beta = 0$ the two oscillators decouple and the gain oscillator grows for any $\gamma>0$: without odd friction there is no unbroken $\mathscr{PT}$ phase at all. Odd friction transfers energy from the amplified to the damped oscillator (at the Rabi rate $\beta/2$, cf. \rr{amp1}) and thereby stabilizes the pair throughout the window
\be
0 < \gamma < \beta \qquad (\textrm{unbroken } \mathscr{PT}),
\ee
in which all four eigenvalues \rr{evs2} are purely imaginary (the identity $(\gamma^2-\beta^2-2\omega^2)^2 - (\gamma^2-\beta^2)(\gamma^2-\beta^2-4\omega^2) = 4\omega^4$ guarantees this). In short: odd friction protects the $\mathscr{PT}$ symmetry. Since the odd force does no work, this protection costs nothing energetically, a design principle for stabilizing driven mechanical and hydrodynamic systems.

\begin{figure}
\begin{center}
\includegraphics[width=3.2in]{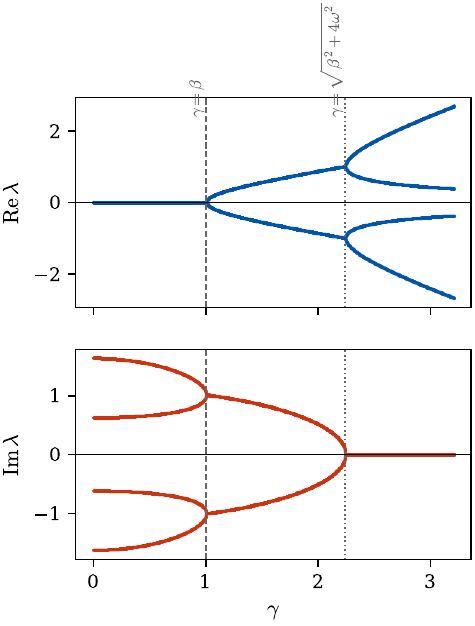}
\end{center}
\vspace{-5pt}
\caption{Real and imaginary parts of the four eigenvalues \rr{evs2} as functions of the loss-gain parameter $\gamma$, for odd friction $\beta = 1$ and $\omega = 1$. For $\gamma\in(0,1)$ all roots are imaginary (unbroken $\mathscr{PT}$; Rabi oscillations at the two frequencies \rr{splitting}). At $\gamma = \beta$ the exceptional point is crossed and for $\gamma\in(1,\sqrt5)$ the roots form two complex-conjugate pairs. For $\gamma\in(\sqrt5,\infty)$ all four roots are real (overdamped). Dashed and dotted vertical lines mark $\gamma = \beta$ and $\gamma = \sqrt{\beta^2+4\omega^2}$, respectively.
\label{fig: eigsgamma} }
\end{figure}

The eigenvalues define three distinct ranges, displayed schematically in Fig. \ref{fig1} and computed in Fig. \ref{fig: eigsgamma}. The system exhibits a twofold transition, in precise analogy with the loss-gain experiments interpreted in \cite{Bender2013,Peng2014}, with the odd friction playing the role of the coupling:
\begin{enumerate}[(i)]
\item
at $\gamma = \beta$ the $\mathscr{PT}$ symmetry breaks at an exceptional point (EP), where the four eigenvalues coalesce pairwise into $\pm i\omega$, each a $2\times2$ Jordan block, and the eigenvectors degenerate;

\item
at $\gamma = \sqrt{\beta^2 + 4\omega^2}$ the broken phase changes character, from two complex-conjugate pairs (oscillatory growth/decay) to four real eigenvalues (overdamped growth/decay). In the unbroken region the motion is in dynamical equilibrium, performing Rabi-type oscillations; the system resembles a closed one although it is in contact with the environment through loss, gain and odd friction.

\end{enumerate}

\begin{figure}
\begin{center}
\includegraphics[width=3.2in]{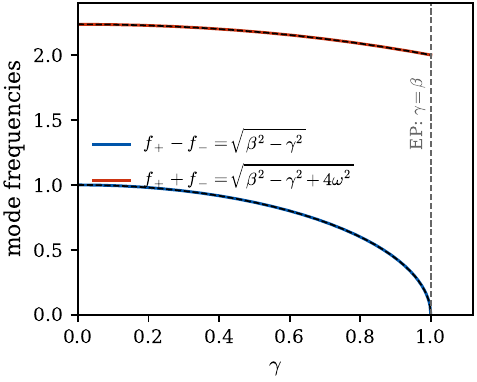}
\end{center}
\vspace{-5pt}
\caption{Exact laws \rr{splitting} for the two Rabi frequencies of the loss-gain odd oscillator in the unbroken phase ($\beta=\omega=1$). The difference $f_+-f_-=\sqrt{\beta^2-\gamma^2}$ (lower curve) collapses with the square-root singularity characteristic of an exceptional point, while the sum (upper curve) remains finite. Dashed curves: the analytic laws.
\label{fig: splitting} }
\end{figure}

In the unbroken phase, writing $\lambda = \pm i f_\pm$ with $f_\pm > 0$, Vieta's formulas applied to \rr{plambda} yield $f_+^2 f_-^2 = \omega^4$ and $f_+^2 + f_-^2 = \beta^2 - \gamma^2 + 2\omega^2$, whence the exact laws
\be \label{splitting}
 f_+ - f_- = \sqrt{\beta^2 - \gamma^2}, \qquad f_+ + f_- = \sqrt{\beta^2 - \gamma^2 + 4\omega^2}, \qquad f_+ f_- = \omega^2.
\ee
The first law generalizes the splitting $f_+-f_-=\beta$ of the pure odd oscillator, Eq. \rr{fpm}, and collapses as $\sqrt{2\beta}\sqrt{\beta - \gamma}$ near the EP: the square-root branching characteristic of exceptional points, which underlies EP-enhanced sensing \cite{Wiersig2014,Chen2017}. A measurement of the two Rabi sidebands as functions of the (tunable) gain therefore determines $\beta$ by extrapolation, with the EP enhancement working in the experimenter's favor; see Fig. \ref{fig: splitting}.

The system \rr{system2a}-\rr{system2b} is derivable from a Lagrangian, and the construction makes the special role of the EP manifest. Write \rr{system2a}-\rr{system2b} as $\ddot{\mathbf{x}} + D\dot{\mathbf{x}} + \omega^2\mathbf{x} = 0$ with $D = \left(\begin{array}{cc} \gamma & -\beta \\ \beta & -\gamma\end{array}\right)$, and seek a symmetric matrix $K$ such that $KD$ is antisymmetric (so that $K\ddot{\mathbf{x}} + KD\dot{\mathbf{x}} + \omega^2 K\mathbf{x} = 0$ is of Euler-Lagrange form). The unique choice, up to scale, is
\be \label{Kmatrix}
K = \begin{pmatrix} 1 & -\gamma/\beta \\ -\gamma/\beta & 1 \end{pmatrix}, \qquad \det K = \frac{\beta^2-\gamma^2}{\beta^2},
\ee
which is singular precisely at the exceptional point $\gamma = \beta$. The Lagrangian and Hamiltonian read
\be \label{LB}
L_2 = \frac{1}{2}\left(\dot x^2 + \dot y^2\right) - \frac{\gamma}{\beta}\,\dot x \dot y + \kappa\left( \dot{x} y - x \dot{y}\right) - \frac{\omega^2}{2}\left(x^2+y^2\right) + \frac{\gamma\omega^2}{\beta}\, x y, \qquad \kappa \equiv \frac{\gamma^2 - \beta^2}{2\beta},
\ee
\be \label{H2}
H_2 = \frac{\beta^2}{2(\beta^2-\gamma^2)}\left[ \pi_x^2 + \pi_y^2 + \frac{2\gamma}{\beta}\,\pi_x \pi_y \right] + \frac{\omega^2}{2}\left(x^2 + y^2\right) - \frac{\gamma\omega^2}{\beta}\,xy ,
\ee
where $\pi_x = p - \kappa y$ and $\pi_y = q + \kappa x$, with $p = \partial L_2/\partial\dot{x} = \dot x - (\gamma/\beta)\dot y + \kappa y$ and $q = \partial L_2/\partial \dot{y} = \dot y - (\gamma/\beta)\dot x - \kappa x$. One verifies directly that Hamilton's equations generated by \rr{H2} reproduce \rr{system2a}-\rr{system2b}; at $\gamma \rightarrow 0$, \rr{LB} and \rr{H2} reduce to \rr{L2} and \rr{H1}. Both $L_2$ and $H_2$ are $\mathscr{PT}$-invariant under \rr{P11}-\rr{T11}, and $H_2$ is a conserved quantity of the loss-gain flow throughout both phases. The construction degenerates at $\gamma = \beta$, where $K$ ceases to be invertible: the breakdown of the Hamiltonian description is not a defect of the method but the canonical signature of the Jordan-block structure at the EP, familiar from Bateman-type constructions \cite{Bateman1931,Dekker1981,Bender2013}. The generalization to arbitrary displacement coupling is given in \ref{sec: genericH}.

\subsection{\label{sec: probe}From odd viscosity to odd friction: the trapped probe}

We now supply the link, promised in the introduction, between the fluid of section \ref{sec: oddNS} and the oscillators of this section. Consider a probe (a disk or colloidal particle of radius $a$ and mass $m$) held by a harmonic trap of stiffness $k_t$ (an optical tweezer, or a magnetic trap) inside a two-dimensional odd viscous liquid, such as the spinning-magnet colloidal liquid of \cite{Soni2019}. In the Stokes regime the force exerted by the liquid on the slowly moving probe is linear in its velocity, and in an odd liquid the resistance tensor acquires an antisymmetric (lift) part alongside the familiar drag \cite{Ganeshan2017,Hosaka2021,Khain2022,Fruchart2023}:
\be \label{probeforce}
\mathbf{F} = -\zeta\, \dot{\mathbf{r}} - \zeta_o\, \epsilon\, \dot{\mathbf{r}}, \qquad \epsilon = \begin{pmatrix} 0 & -1 \\ 1 & 0\end{pmatrix},
\ee
where $\zeta \propto \eta$ is the usual drag coefficient and $\zeta_o \propto \eta_o$ is the odd (transverse) drag; for a disk both coefficients carry the weak logarithmic size dependence characteristic of two-dimensional Stokes flow \cite{Hosaka2021}. Newton's law for the trapped probe then reads
\be \label{probeEOM}
m\ddot{\mathbf{r}} + \zeta \dot{\mathbf{r}} + \zeta_o \epsilon\, \dot{\mathbf{r}} + k_t \mathbf{r} = \mathbf{0},
\ee
which is exactly the odd oscillator \rr{system0} (supplemented by ordinary friction) with
\be \label{betamap}
\beta = \frac{\zeta_o}{m} \propto \frac{\eta_o}{m}, \qquad \omega^2 = \frac{k_t}{m}.
\ee
Two measurement protocols result from the above discussion:
\begin{enumerate}[(i)]
\item \emph{Passive spectroscopy:} by \rr{fpm}, the power spectrum of the probe displacement exhibits two peaks split exactly by $f_+ - f_- = \beta$ (ordinary friction $\zeta$ broadens the peaks without shifting the splitting at leading order), so the peak splitting is a direct mechanical measurement of $\zeta_o$ and hence of $\eta_o$.

\item \emph{Loss-gain thresholds:} endowing one axis of the trap with tunable feedback gain $\gamma$ (routinely implemented in optical-tweezer experiments) and the other with matched loss realizes \rr{system2a}-\rr{system2b}; the collapse of the sidebands according to the exact law $f_+-f_-=\sqrt{\beta^2-\gamma^2}$, and the location $\gamma = \beta$ of the instability threshold, determine $\beta$ redundantly, in the manner of the whispering-gallery experiments \cite{Peng2014} but in a fluid-mechanical setting.

\end{enumerate}

Protocol (ii) benefits from the square-root EP enhancement discussed below \rr{splitting}.

\subsection{\label{sec: stochastic}Stochastic odd oscillator: what thermal noise can and cannot see}

In any experiment of the type just described the probe is subject to thermal (or active) noise, so we consider the Langevin extension of \rr{probeEOM} (set $m=1$),
\be \label{langevin}
\ddot{x} + \gamma_d \dot{x} - \beta\dot y + \omega^2 x = \xi_x, \qquad \ddot{y} + \gamma_d \dot{y} + \beta\dot x + \omega^2 y = \xi_y,
\ee
with isotropic equilibrium noise, $\langle \xi_i(t)\xi_j(t')\rangle = 2\gamma_d T\, \delta_{ij}\delta(t-t')$. Solving the stationary Lyapunov problem for \rr{langevin} one finds that all equal-time cross-correlations vanish,
\be \label{BvL}
\langle x y\rangle = \langle x\dot y\rangle = \langle y \dot x\rangle = \langle x\dot y - y\dot x\rangle = 0, \qquad \langle x^2\rangle = \langle y^2 \rangle = T/\omega^2,
\ee
independently of $\beta$: this is the Bohr-van Leeuwen theorem in oscillator form (the odd force is Lorentz-like, so the Boltzmann distribution does not feel it). Static statistics are therefore blind to odd friction. The dynamics is not: the response to a small applied force $\mathbf{f}e^{-i\varpi t}$ is $\hat{\mathbf{r}} = \chi(\varpi)\hat{\mathbf{f}}$ with
\be \label{response}
\chi(\varpi) = \frac{1}{D_0^2 - \varpi^2\beta^2}\begin{pmatrix} D_0 & -i\varpi\beta \\ i\varpi\beta & D_0 \end{pmatrix}, \qquad D_0 = \omega^2 - \varpi^2 - i\varpi\gamma_d,
\ee
whose antisymmetric part is proportional to $\beta$, and the displacement cross-spectrum inherits it:
\be \label{crossspec}
S_{xy}(\varpi) = 2\gamma_d T\, \left[\chi(\varpi)\chi^\dagger(\varpi)\right]_{xy} = \frac{-4i\,\gamma_d T\, \beta\, \varpi \left(\omega^2 - \varpi^2\right)}{\left| D_0^2 - \varpi^2\beta^2 \right|^2}.
\ee
$S_{xy}$ is purely imaginary (consistent with \rr{BvL}: the equal-time correlation, its frequency integral, vanishes), odd in $\varpi$, and proportional to $\beta$: time-resolved passive microrheology of the trapped probe measures the odd friction, and hence $\eta_o$, with no applied drive at all. In an active chiral, however, the noise itself violates the fluctuation-dissipation relation and acquires odd correlations \cite{Han2021,Hargus2021}; comparing the measured $S_{xy}$ with the equilibrium form \rr{crossspec} then quantifies the departure. The autospectra $S_{xx}=S_{yy}$ display the two peaks at $f_\pm$ whose splitting realizes protocol (i) of section \ref{sec: probe}.

\subsection{\label{sec: nonlinear}Nonlinear saturation of the broken phase}

\begin{figure}
\begin{center}
\includegraphics[width=6.4in]{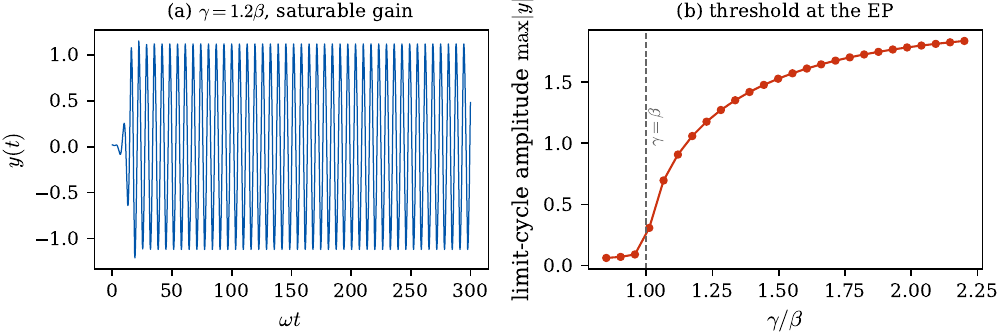}
\end{center}
\vspace{-5pt}
\caption{Saturation of the broken $\mathscr{PT}$ phase by saturable gain, Eq. \rr{vdp} with $\beta=\omega=y_s=1$. (a) Above the exceptional point ($\gamma = 1.2\beta$) the amplitude grows and settles onto a stable limit cycle. (b) Limit-cycle amplitude versus $\gamma/\beta$: a continuous, lasing-like threshold located at the exceptional point $\gamma=\beta$ (dashed line).
\label{fig: nonlinear} }
\end{figure}

Beyond the exceptional point the linear theory predicts unbounded growth, so the fate of the broken phase is decided by nonlinearity. Two qualitatively different nonlinearities must be distinguished. A conservative anharmonicity (e.g. a Duffing term $\delta x^3$, $\delta y^3$ added to \rr{system2a}-\rr{system2b}) detunes the two oscillators but extracts no energy; since the gain $-\gamma\dot y$ injects energy at all amplitudes, growth continues---our numerical integrations show no saturation, only a nonlinear modification of the growth. In contrast, a saturable gain, which is what any physical gain provides, cf. lasing systems \cite{Peng2014,ElGanainy2018}, of van der Pol form,
\be \label{vdp}
 \ddot{x} + \gamma \dot{x} - \beta \dot{y} + \omega^2 x = 0, \qquad
 \ddot{y}  - \gamma\left(1 - y^2/y_s^2\right)\dot{y}+ \beta \dot{x} + \omega^2 y = 0,
\ee
arrests the growth: the broken phase flows to a stable limit cycle whose amplitude vanishes as $\gamma \rightarrow \beta^+$ and grows continuously beyond it (Fig. \ref{fig: nonlinear}). The $\mathscr{PT}$ transition of the odd loss-gain pair is thus observable as a soft, lasing-like threshold at $\gamma = \beta$: below threshold, noise-broadened Rabi sidebands obeying \rr{splitting}; above threshold, a finite-amplitude limit cycle. This is precisely the phenomenology familiar from $\mathscr{PT}$-symmetric photonics \cite{Peng2014,ElGanainy2018}, now predicted for a mechanical probe in an odd viscous liquid.

\section{\label{sec: quantum}Quantum odd oscillators and quantum odd fluids}

Odd viscosity originated in the quantum Hall effect \cite{Avron1995}, and the correspondence \rr{hbar} suggests that the natural home of the structures derived above is quantum mechanical. In this section we quantize the odd oscillator (section \ref{sec: fockdarwin}), formulate the loss-gain pair as a quantum master equation and locate its Liouvillian exceptional point (section \ref{sec: lindblad}), and outline the contact with quantum Hall fluids (section \ref{sec: QH}), graphene electron hydrodynamics (section \ref{sec: graphene}), chiral superfluids (section \ref{sec: chiralSF}) and synthetic platforms (section \ref{sec: synthetic}).

\subsection{\label{sec: fockdarwin}Quantization of the odd oscillator: the Fock-Darwin spectrum}

The Hamiltonian \rr{H1} is real and quadratic, hence directly quantizable: $[\hat x, \hat p] = [\hat y, \hat q] = i\hbar$. By the magnetic analogy noted below \rr{fpm}, $\hat H_1$ is the Fock-Darwin Hamiltonian \cite{Fock1928,Darwin1930} of a charged particle in a uniform magnetic field of cyclotron frequency $\beta$ in an isotropic trap $\omega$, and its exact spectrum is
\be \label{FDspectrum}
E_{n_+ n_-} = \hbar\, \omega_+ \left( n_+ + \tfrac12 \right) + \hbar\, \omega_- \left( n_- + \tfrac12 \right), \qquad \omega_\pm = \sqrt{\omega^2 + \frac{\beta^2}{4}} \pm \frac{\beta}{2}, \qquad n_\pm = 0, 1, 2, \ldots
\ee
with angular momentum $L_z = \hbar(n_- - n_+)$. The mode frequencies $\omega_\pm$ are exactly the classical frequencies $f_\pm$ of Eq. \rr{fpm}: the classical $\mathscr{PT}$ analysis lifts to the quantum spectrum without modification. The quantum $\mathscr{PT}$ operation is implemented by the unitary $\hat{\mathscr{P}}$ realizing \rr{P11} together with the antiunitary $\hat{\mathscr{T}}$ realizing \rr{T11}, and $[\hat H_1, \hat{\mathscr{P}}\hat{\mathscr{T}}] = 0$ while $\hat H_1$ commutes with neither factor separately. Quantization is performed with the representative \rr{H1}, singled out
among the equivalent alternatives of \ref{sec: nullapp} by minimal
coupling and by the conserved energy $E_1$; $s$-equivalent classical
Lagrangians can in general quantize inequivalently \cite{Morandi1990}.

\begin{figure}
\begin{center}
\includegraphics[width=3.2in]{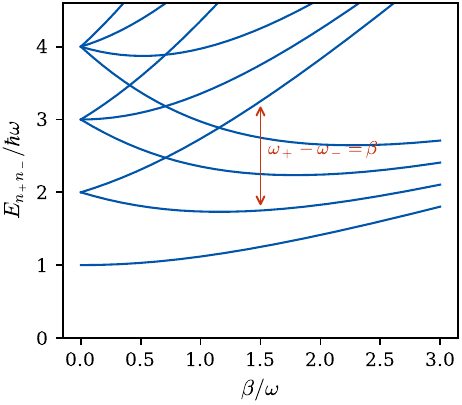}
\end{center}
\vspace{-5pt}
\caption{Fock-Darwin spectrum \rr{FDspectrum} of the quantized odd oscillator versus odd friction $\beta$ (levels with $E<4.6\,\hbar\omega$ shown). The splitting between the levels $(n_+,n_-) = (1,0)$ and $(0,1)$ (arrow, drawn at $\beta = 1.5\,\omega$) equals $\hbar\beta$ exactly, Eq. \rr{qsplit}. As $\beta/\omega \rightarrow \infty$ the levels reorganize into Landau levels of spacing $\hbar\beta$.
\label{fig: fockdarwin} }
\end{figure}

Three consequences deserve emphasis. First, the level splitting
\be \label{qsplit}
\omega_+ - \omega_- = \beta
\ee
is exact at all $\beta$ (Fig. \ref{fig: fockdarwin}): spectroscopy of the quantized odd oscillator measures the odd friction directly; this is the quantum counterpart of protocol (i) in section \ref{sec: probe}. Second, in the limit $\omega \rightarrow 0$ the spectrum \rr{FDspectrum} collapses to Landau levels with spacing $\hbar\beta$: the odd oscillator interpolates continuously between a trapped degree of freedom and a quantum Hall problem, closing the conceptual loop with the origin of odd viscosity \cite{Avron1995,Read2009}. Third, the correspondence is not merely formal: Fock-Darwin spectra are routinely measured in semiconductor quantum dots \cite{Tarucha1996}, in Penning traps, and in rotating Bose-Einstein condensates, where artificial gauge fields implement precisely a tunable $\beta$ for neutral atoms \cite{Dalibard2011}. Any of these platforms is, in the present language, a quantum odd oscillator.

\subsection{\label{sec: lindblad}The open quantum pair: Liouvillian exceptional point}

To treat loss and gain quantum mechanically we first reduce \rr{system2a}-\rr{system2b} to slowly varying amplitudes (rotating-wave approximation, RWA, valid for $\beta,\gamma \ll \omega$): with $x = Ae^{i\omega t} + c.c.$, $y = Be^{i\omega t}+c.c.$,
\be \label{RWA}
\dot A = -\frac{\gamma}{2} A + \frac{\beta}{2} B, \qquad \dot B = +\frac{\gamma}{2}B - \frac{\beta}{2} A ,
\ee
with eigenvalues $\pm\frac12\sqrt{\gamma^2 - \beta^2}$: the RWA retains the $\mathscr{PT}$ transition at $\gamma = \beta$ and the unbroken-phase splitting $\sqrt{\beta^2-\gamma^2}$, in agreement with the exact law \rr{splitting} (the second transition at $\gamma = \sqrt{\beta^2+4\omega^2}$ is a beyond-RWA effect, invisible at this order). Equations \rr{RWA} are the first-moment equations of the quantum master equation
\be \label{lindblad}
\dot{\hat\rho} = -\frac{i}{\hbar}[\hat H_0 + \hat H_c, \hat\rho] + \gamma\, \mathcal{D}[\hat a]\hat\rho + \gamma\, \mathcal{D}[\hat b^\dagger]\hat\rho, \qquad \hat H_c = i\hbar\frac{\beta}{2}\left( \hat a^\dagger \hat b - \hat a \hat b^\dagger \right),
\ee
where $\hat H_0 = \hbar\omega(\hat a^\dagger \hat a + \hat b^\dagger \hat b)$, $\mathcal{D}[\hat L]\hat\rho = \hat L \hat\rho \hat L^\dagger - \frac12\{\hat L^\dagger \hat L, \hat\rho\}$, the $\hat a$ mode is lossy, and the $\hat b$ mode is linearly amplified. The coupling $\hat H_c$---an excitation-conserving beam-splitter coupling with a $\pi/2$ phase---is the quantum image of odd friction.

\begin{figure}
\begin{center}
\includegraphics[width=3.2in]{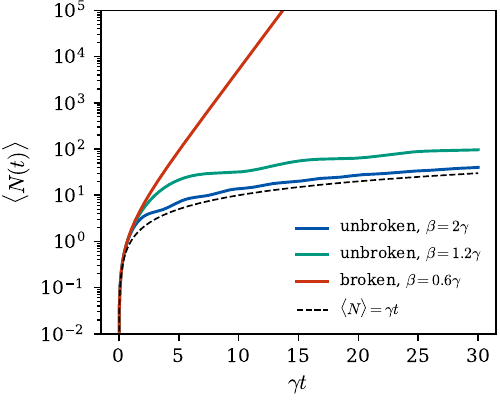}
\end{center}
\vspace{-5pt}
\caption{Growth of the total occupation $\langle N(t)\rangle$ of the quantum loss-gain pair \rr{lindblad}, from the moment equations \rr{moments} with vacuum initial conditions. In the unbroken phase ($\beta > \gamma$) the growth is asymptotically linear with universal slope $\gamma$ (dashed line), modulated by Rabi oscillations at $\sqrt{\beta^2 - \gamma^2}$; in the broken phase ($\beta<\gamma$) it is exponential at rate $\sqrt{\gamma^2-\beta^2}$. The Liouvillian exceptional point $\beta = \gamma$ separates the two heating regimes.
\label{fig: lindblad} }
\end{figure}

First moments $\langle \hat a\rangle, \langle\hat b\rangle$ obey exactly \rr{RWA} and reproduce the classical twofold phenomenology; the exceptional point of the Liouvillian coincides here with the mean-field EP, as is generic for quadratic Liouvillians \cite{Minganti2019}. The quantum content appears at second order. The occupations $n_a = \langle\hat a^\dagger \hat a\rangle$, $n_b = \langle \hat b^\dagger \hat b\rangle$ and coherence $s = \langle \hat a^\dagger \hat b + \hat b^\dagger \hat a\rangle$ close on themselves (\ref{sec: lindbladapp}):
\be \label{moments}
\dot n_a = -\gamma n_a + \frac{\beta}{2} s, \qquad \dot n_b = +\gamma\left( n_b + 1 \right) - \frac{\beta}{2}s, \qquad \dot s = \beta\left( n_b - n_a \right),
\ee
where the inhomogeneous term $+\gamma$ is spontaneous emission into the gain mode: amplified vacuum noise. The eigenvalues of the homogeneous part of \rr{moments} are $\{0, \pm\sqrt{\gamma^2-\beta^2}\}$, and the zero mode carries the inhomogeneity, with the exact consequence (\ref{sec: lindbladapp})
\be \label{heating}
\langle N(t)\rangle \equiv n_a + n_b \simeq \gamma\, t \quad (\gamma < \beta, \textrm{ unbroken}), \qquad
\langle N(t) \rangle \sim e^{\sqrt{\gamma^2 - \beta^2}\, t} \quad (\gamma > \beta, \textrm{ broken}),
\ee
see Fig. \ref{fig: lindblad}. Thus, although no quantum system with linear gain is stationary, the quantum-optical no-go of \cite{Scheel2018}, the Liouvillian EP separates two measurably distinct heating regimes: linear growth with universal slope $\gamma$ (independent of $\beta$), decorated by coherent oscillations at $\sqrt{\beta^2-\gamma^2}$ which measure the odd coupling, versus exponential growth at rate $\sqrt{\gamma^2-\beta^2}$. Superconducting circuits, optomechanical systems and the photonic platforms of \cite{Peng2014,ElGanainy2018} realize \rr{lindblad} directly, with the beam-splitter phase of $\hat H_c$ implementing the odd coupling; the photon-number transient \rr{heating} is the experimental signature.

\subsection{\label{sec: QH}Quantized odd viscosity in quantum Hall fluids}

In a gapped quantum Hall state the odd (Hall) viscosity is quantized, $\eta_H = \frac{\hbar}{2} \bar{s}\, n$, with $n$ the particle density and $\bar s$ the mean orbital spin per particle \cite{Avron1995,Read2009,ReadRezayi2011}; it is related by Ward identities to the wave-vector dispersion of the Hall conductivity, $\sigma_H(q) = \sigma_H(0)\left[1 + C (q\ell)^2 + \ldots\right]$ with the coefficient $C$ determined by $\eta_H$ and $\ell$ the magnetic length \cite{HoyosSon2012,Bradlyn2012,Hoyos2014}. The long-wavelength dynamics of such states is governed by hydrodynamic equations of precisely the odd Navier-Stokes form of section \ref{sec: oddNS} (with $2\Omega \rightarrow \omega_c$, the cyclotron frequency), so the $\mathscr{PT}$ statements proved there---including the invariance of the maximal-stress observable \rr{shearmax}---carry over verbatim to quantum Hall hydrodynamics. In this setting the point-pair symmetry below \rr{shearmax} becomes a symmetry between strain-response measurements at mirror-related positions, complementary to the established $\sigma_H(q)$ route to $\eta_H$.

\subsection{\label{sec: graphene}Viscous electron fluids in graphene}

The one existing measurement of odd viscosity in a quantum system is hydrodynamic: the Hall viscosity of graphene's electron fluid was extracted from magnetotransport in the hydrodynamic regime \cite{Berdyugin2019}, following the proposal of \cite{Pellegrino2017}; see \cite{Lucas2018} for a review of electron hydrodynamics. The governing equations are \rr{onsx}-\rr{onsy} with the Coriolis frequency replaced by the cyclotron frequency and with momentum-relaxing friction added, i.e. exactly the setting of this paper. Two consequences follow.
\begin{enumerate}[(i)]
\item
The $\mathscr{PT}$ invariance of section \ref{sec: oddNS} implies a symmetry of nonlocal transport: with current injected and drained at contacts placed at mirror-related positions ($x\leftrightarrow y$ exchange accompanied by field reversal, cf. \rr{P1}-\rr{T1}), the vicinity resistances measured in the two configurations coincide, and the antisymmetric deviation isolates the $\eta_o$ contribution---a transport analogue of the stress protocol of section \ref{sec: maxstress}. 
\item 
A gate-defined electrostatic trap for an electron puddle in the hydrodynamic regime realizes the trapped-probe system of section \ref{sec: probe} with $\beta$ set by $\eta_o$ and $\omega_c$; its collective dipole mode is the odd oscillator, and the splitting law \rr{fpm} governs its magneto-spectroscopy (in the Galilean limit, Kohn's theorem fixes the bare dipole frequencies, and deviations track the viscous contribution).
\end{enumerate}

\subsection{\label{sec: chiralSF}Chiral superfluids and superconductors}

Odd viscosity does not require an external magnetic field: it arises from spontaneous breaking of time reversal, as in thin films of $^3$He-A and in candidate $p_x + ip_y$ superconductors, where it is again tied to the orbital spin of the pairs \cite{Read2009,ReadRezayi2011,Hoyos2014}. These systems realize the ``spontaneous'' clause of the opening sentence of this paper: the $\mathscr{PT}$ symmetry \rr{P1}-\rr{T1} is then an emergent symmetry of the order-parameter hydrodynamics, and the trapped-probe and wave protocols of sections \ref{sec: probe} and \ref{sec: hydroPT} apply with $\eta_o$ of intrinsic, quantum origin. We leave the detailed adaptation---in particular the role of the gapless edge in a bounded sample \cite{Abanov2018}---to future work.

\subsection{\label{sec: synthetic}Synthetic gauge fields and photonic Landau levels}

Finally, engineered platforms provide odd oscillators with complete parameter control. Artificial gauge potentials give neutral cold atoms a tunable $\beta$ \cite{Dalibard2011}, so a harmonically trapped gas realizes \rr{FDspectrum} directly. Twisted optical resonators realize Landau levels for photons \cite{Schine2016}; since photon loss is intrinsic and gain is supplied by pumping, these systems are natural hosts for the loss-gain master equation \rr{lindblad}, with the heating dichotomy \rr{heating} as the smoking gun of the Liouvillian exceptional point. Gyroscopic metamaterials \cite{Nash2015} provide the classical mechanical analogue with $\beta$ set by the spinning speed, where the protection law $\gamma_{PT} = \beta$ of section \ref{sec: frictionlossgain} can be tested by motorized gain.

\section{\label{sec: conclusions}Conclusions}
In this paper we established consequences of $\mathscr{P\!T}$-symmetry, in the sense of Bender \cite{Bender2019}, for odd viscous liquids and their oscillator and quantum counterparts.

For the liquid: the odd Navier-Stokes equations of a two-dimensional compressible liquid or a three-dimensional incompressible liquid, including Coriolis forces, are invariant under the combined $\mathscr{PT}$ symmetry; the odd terms derive from a Lagrangian density whose conserved Hamiltonian expresses the work-free nature of odd viscosity and whose Noether momentum density is the Schr\"odinger current of the complex velocity $\psi = u + iv$, consistent with the fact that the linearized bulk equations are a Schr\"odinger equation with $\nu_o \leftrightarrow \hbar/2m$; Ertel's construction extends to odd liquids in the form of a local conservation law for potential vorticity; and transverse waves subject to gain and loss undergo a $\mathscr{PT}$ transition at the odd-viscosity-determined wavenumber $k_\star = \sqrt{\gamma/\nu_4}$.

For the oscillators: a loss-gain pair coupled by odd friction exhibits a twofold transition---$\mathscr{PT}$ breaking at the exceptional point $\gamma=\beta$, followed by an oscillatory-to-overdamped transition at $\gamma = \sqrt{\beta^2+4\omega^2}$---with the exact laws $f_+ - f_- = \sqrt{\beta^2-\gamma^2}$, $f_+ + f_- = \sqrt{\beta^2-\gamma^2+4\omega^2}$ for the Rabi frequencies. Odd friction protects $\mathscr{PT}$ symmetry: without it the unbroken phase does not exist. The pair possesses a conserved $\mathscr{PT}$-symmetric Hamiltonian \rr{H2} that degenerates precisely at the exceptional point. A probe trapped in an odd viscous liquid realizes this system with $\beta = \zeta_o/m \propto \eta_o$, yielding two mechanical measurement protocols for odd viscosity (spectral splitting; loss-gain thresholds), a stochastic signature (the cross-spectrum \rr{crossspec}, which survives the Bohr-van Leeuwen blindness of static equilibrium statistics), and a lasing-like saturation of the broken phase under saturable gain.

For the quantum systems: the quantized odd oscillator has the Fock-Darwin spectrum with level splitting exactly $\hbar\beta$, interpolating to Landau levels as $\omega\rightarrow 0$; the open loss-gain pair, formulated as a quadratic Lindblad master equation, has a Liouvillian exceptional point at $\gamma = \beta$ separating linear heating with universal slope $\gamma$ from exponential heating at rate $\sqrt{\gamma^2-\beta^2}$; and the framework connects directly to the quantized Hall viscosity of quantum Hall fluids, to the measured Hall viscosity of graphene's electron fluid, to chiral superfluids, and to synthetic gauge-field and photonic platforms.

We believe these results elevate the $\mathscr{PT}$ perspective on odd transport from an observation about symmetries to a set of concrete, requisite experimental protocols for measuring odd viscosity coefficients in classical and quantum fluids alike.

\section*{Acknowledgements}

The authors are grateful to Carl Bender for helpful discussions during meetings at the University of Washington and Northwestern University, that partially motivated this work. The work of A. L. was supported by NSF Grant No. DMR-2452658 and H. I. Romnes Faculty Fellowship provided by the University of Wisconsin-Madison Office of the Vice Chancellor for Research and Graduate Education with funding from the Wisconsin Alumni Research Foundation. The authors acknowledge the use of Claude (Anthropic) \cite{Claude2026} with manuscript preparation, which includes in particular numerics, graphics, and symbolic verification of analytics. All results were conceptualized, checked, and validated by the authors.  

\section*{Data availability}

All data presented in the figures were generated from analytical expressions derived and defined in the paper. The code used to produce the plots will be made available by the authors upon reasonable request.

\appendix

\section{\label{sec: ErtelApp}Simplification of the potential-vorticity flux}
We show here that $\mathbf{F}\cdot \nabla \lambda$ simplifies and gives rise to the law \rr{cfluidvol}.
This is achieved by realizing that the odd and shear viscous terms in the flux $\mathbf{F}$ in Eq. \rr{cons1} can be replaced by equivalent expressions having the same divergence. 
\be \label{Fcdot}
\mathbf{F}\cdot \nabla \lambda = q\mathbf{v}\cdot \nabla\lambda - q \frac{D \lambda}{Dt}  - \lambda \partial_z \mathcal{S} (\hat{\mathbf{x}}u + \hat{
\mathbf{y}} v)\cdot \nabla \lambda + \nu \nabla^2 \mathbf{v} \times \nabla \lambda \cdot \nabla \lambda.
\ee
Above, the mixed product arises by considering the identity  $\textrm{div} (\nu \nabla^2 \mathbf{v} \times \nabla \lambda)  = \textrm{div} (\lambda  \nu \nabla^2  \textrm{curl}\mathbf{v}), $ which thus vanishes. 

The odd term $\lambda \partial_z \mathcal{S} (\hat{\mathbf{x}}u + \hat{
\mathbf{y}} v)$ appearing in the flux of \rr{cons1} can be replaced by
\be
 \mathbf{G} \times \nabla \lambda,
\ee
where $\mathbf{G} = \mathcal{S} (v,-u,0)$. This is the case because
$\nabla \cdot ( \mathbf{G} \times \nabla \lambda ) = \nabla \lambda \cdot \nabla \times \mathbf{G} = \nabla \lambda\cdot \partial_z \mathcal{S} (\hat{\mathbf{x}}u + \hat{
\mathbf{y}} v)$ and
$\nabla \cdot (\lambda \nabla \times \mathbf{G} ) = \nabla \lambda \cdot \nabla \times \mathbf{G}$ and
$\nabla \times \mathbf{G} =  \partial_z \mathcal{S} (\hat{\mathbf{x}}u + \hat{
\mathbf{y}} v)$, where we used the incompressibility condition. Thus, the second to last term in \rr{Fcdot} vanishes.

Finally, using the identity 
$\frac{D \lambda}{Dt}  = \partial_t \lambda + \mathbf{v} \cdot \nabla \lambda$, the first two terms in the right hand-side of \rr{Fcdot} give $q \partial_t\lambda$.
This discussion then leads to the law \rr{cfluidvol}.

\section{\label{sec: 3D}Three-dimensional odd constitutive law}
The three-dimensional odd constitutive law \citep[\S 13]{Landau1981}
can be expressed in Cartesian coordinates as
\be \label{sigma3}
{\sigma} = \eta_o 
\left(\begin{array}{ccc}
-\left(\partial_x v + \partial_y u \right) &  \partial_x u  - \partial_y v & 0\\
 \partial_x u  - \partial_y v & \partial_x v + \partial_y u   & 0\\
0&0&0
\end{array}
\right)
+ \eta_4 
\left(\begin{array}{ccc}
0 & 0  & -(\partial_y w + \partial_z v)\\
0 & 0  & \partial_x w + \partial_z u\\
 -(\partial_y w + \partial_z v)&\partial_x w + \partial_z u&0
\end{array}
\right). 
\ee
The simple form \rr{NS2} of the Navier-Stokes equations arises after some algebra by introducing the modified pressure $\tilde{p} = p + \eta_4 (\partial_x v - \partial_yu)$ and employing the incompressibility condition $u_x + v_y + w_z =0$.

\section{\label{sec: nullapp}Non-uniqueness of the inertial generating term
and the Helmholtz test}

We make precise the freedom in the term $\frac12(u^2v_x - v^2u_y)$ of
\rr{Ldensity}. Consider the most general density cubic in the velocities
and linear in their first spatial gradients,
\be \label{Ngen}
\mathcal{N} = \sum_{f\,\in\,\{u^2,\,v^2,\,uv\}}\;
\sum_{g\,\in\,\{u_x,\,u_y,\,v_x,\,v_y\}} a_{fg}\, f\, g,
\ee
a twelve-parameter family, and require that the Euler-Lagrange equations of
$\frac12(uv_t - vu_t) + \mathcal{N}$ reproduce the inertial terms of
\rr{onsx}-\rr{onsy} (at $\Omega=0$), allowing in addition for a
redefinition of the pressure by an arbitrary cubic $\chi(u,v)$. Direct
computation shows that the solution set is the eight-parameter affine
family
\be \label{Nfamily}
\mathcal{N} = \frac{1}{2}\left(u^2 v_x - v^2 u_y\right)
+ D_x G(u,v) + D_y H(u,v),
\ee
with $G$ and $H$ arbitrary homogeneous cubics and $D_x, D_y$ total
derivatives, while the pressure redefinition is forced to vanish,
$\chi \equiv 0$. The freedom in \rr{Nfamily} is thus exactly the space of
null Lagrangians: every admissible representative differs from every other
by an exact divergence (for instance $\frac12 u^2 v_x$ may be traded for
$-uv\,u_x$ using $G = u^2 v/2$) and yields identical, not merely
equivalent, field equations. Because the null terms contain no time
derivatives, the canonical momenta $p = -v/2$, $q = u/2$ and the symplectic
structure are unchanged; the Hamiltonian and momentum densities shift by
spatial divergences, so the charges are representative-independent under
decaying or periodic boundary conditions, while densities and boundary
fluxes are not. On bounded domains \cite{Abanov2018}, or under coupling to
external fields, this residual scheme dependence must be fixed by a stated
choice of representative.

The existence of the principle itself is explained by the Helmholtz
self-adjointness criterion of the inverse problem of the calculus of
variations \cite{Olver1993,Morandi1990}, in view of the classical no-go
results for Eulerian variational formulations of viscous flow
\cite{Millikan1929,Finlayson1972}. The kinetic term $\frac12(uv_t-vu_t)$
pairs the two velocity components symplectically, casting the equations in
the form $\epsilon\,\partial_t\bm{\phi} = \delta\mathcal{F}/\delta\bm{\phi}$
for a functional $\mathcal{F}$, with $\bm{\phi}=(u,v)^T$ and $\epsilon$ the
antisymmetric unit matrix. A variational derivative necessarily possesses a
symmetric kernel. The odd viscous terms derive from
$\int \frac{\eta_o}{2}|\nabla\bm{\phi}|^2$, whose variational derivative
$-\eta_o\nabla^2\bm{\phi}$ is symmetric: odd viscosity is variational in
this pairing. Shear viscosity would instead require
$\delta\mathcal{F}/\delta\bm{\phi} = -\nu\,\epsilon\,\nabla^2\bm{\phi}$, an
antisymmetric kernel that no functional can produce: even viscosity is not
variational within this pairing, for any choice of representative, and its
inclusion demands Bateman's doubling of the degrees of freedom
\cite{Bateman1931,Morse1953}. Genuinely distinct freedoms lie outside the
class \rr{Ngen}: multiplication of the equations by constant invertible
matrices produces ``$s$-equivalent'' Lagrangians not related by null terms
\cite{Morandi1990}---the finite-dimensional counterpart is the $K$-matrix
construction of section \ref{sec: frictionlossgain}---while changes of
field space (Clebsch potentials) underlie the classical escape routes
\cite{SeligerWhitham1968,Salmon1988} as well as the free-surface
variational principle for incompressible odd flow of
Ref.~\cite{AbanovMonteiro2019}. We note finally that the three-dimensional
density \rr{Ldensity3} is claimed only for the odd viscous terms: the
three-dimensional inertial terms in Eulerian velocity variables encounter
the classical Lin-constraint obstruction \cite{Salmon1988} and are not
reproduced by it.

\section{\label{sec: genericH}Hamiltonian of the generic odd loss-gain system}
For the generic system
\be \label{genericB}
\ddot{\mathbf{x}} +  \left( \begin{array}{cc}
 \gamma & -\beta \\
 \beta & - \gamma
 \end{array} \right) \dot{\mathbf{x}} + \left( \begin{array}{cc}
 \omega^2 & g \\
 g & \omega^2
 \end{array} \right) \mathbf{x} = \mathbf{0}
\ee
the construction of section \ref{sec: frictionlossgain} applies without modification: with $D$ and $W$ the two matrices in \rr{genericB}, the matrix $K$ of \rr{Kmatrix} renders $KD$ antisymmetric, and $KW$ is automatically symmetric for any $g$. The Lagrangian is
\be \label{LBgen}
L = \frac{1}{2}\left(\dot x^2 + \dot y^2\right) - \frac{\gamma}{\beta}\dot x\dot y + \kappa\left(\dot x y - x \dot y\right) - \frac{1}{2}\left(\omega^2 - \frac{\gamma g}{\beta}\right)\left(x^2 + y^2\right) - \left( g - \frac{\gamma\omega^2}{\beta}\right) xy,
\ee
with $\kappa = (\gamma^2-\beta^2)/2\beta$ as in \rr{LB}, and the Hamiltonian
\be \label{H2gen}
H = \frac{\beta^2}{2(\beta^2 - \gamma^2)}\left[ \pi_x^2 + \pi_y^2 + \frac{2\gamma}{\beta}\pi_x\pi_y\right] + \frac{1}{2}\left( \omega^2 - \frac{\gamma g}{\beta}\right)\left(x^2+y^2\right) + \left( g - \frac{\gamma \omega^2}{\beta}\right) xy,
\ee
with $\pi_x = p - \kappa y$, $\pi_y = q + \kappa x$. We verified by direct computation that Hamilton's equations generated by \rr{H2gen} reproduce \rr{genericB}. As in the main text, $\det K = (\beta^2-\gamma^2)/\beta^2$ vanishes at the exceptional point $\gamma = \pm\beta$, where the Hamiltonian description degenerates; and at $\gamma \rightarrow 0$, $g \rightarrow 0$ one recovers \rr{H1}. The system \rr{genericB} interpolates between the position-coupled loss-gain oscillators of Bender et al. \cite{Bender2013} ($\beta \rightarrow 0$, where a Bateman-type Hamiltonian with cross kinetic term $\propto \dot x\dot y$ survives after rescaling $K$ by $\beta$) and the velocity-coupled odd oscillator studied here ($g = 0$).

\section{\label{sec: freqapp}Frequency interpretation of the eigenvalues}
\begin{figure}
\begin{center}
\includegraphics[height=1.2in,width=6in]{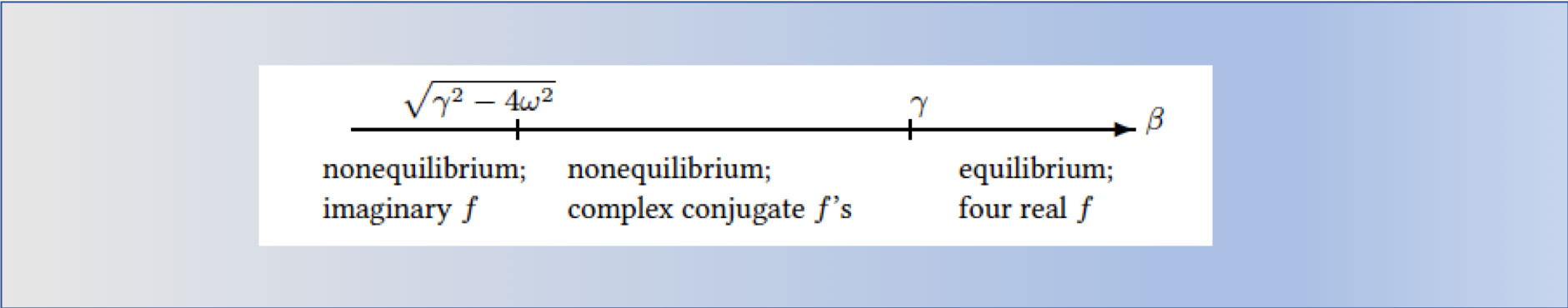}
\end{center}
\vspace{-1pt}
\caption{Regimes that determine the character of the frequencies \rr{evs3} as the odd friction (coupling) $\beta$ is varied at fixed $\gamma > 2\omega$.
In the left-most region the frequencies \rr{evs3} are imaginary; in the middle region they become two complex conjugate pairs;
in the right-most region ($\beta>\gamma$, unbroken $\mathscr{PT}$) all four are real. These three regimes can be seen in Fig. \ref{fig: freqsbeta}.
\label{fig2} }
\vspace{-10pt}
\end{figure}

\begin{figure}
\begin{center}
\includegraphics[width=3.2in]{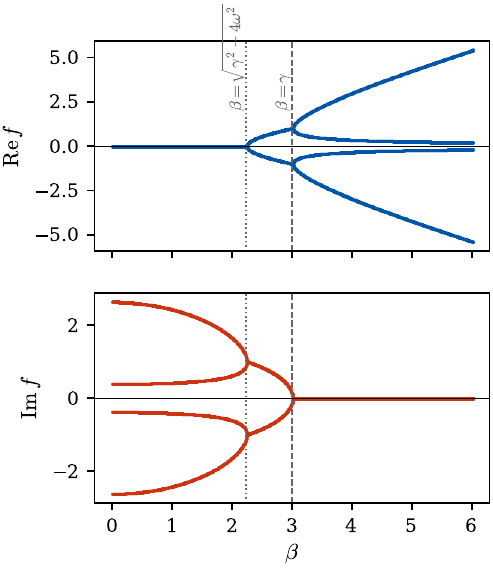}
\end{center}
\vspace{-5pt}
\caption{Real (top) and imaginary (bottom) parts of the frequencies \rr{evs3} versus the coupling (odd friction) $\beta$, at fixed $\gamma=3$ and $\omega=1$; compare with the experiments referenced in \cite{Bender2013,Peng2014}. Dotted and dashed vertical lines mark $\beta = \sqrt{\gamma^2-4\omega^2}$ and $\beta = \gamma$, respectively.
\label{fig: freqsbeta} }
\end{figure}

With a view to comparison with the whispering-gallery experiments
of Bender and coworkers \cite{Peng2014} and their theoretical interpretation \cite{Bender2013},
we substitute $\lambda = if$ in \rr{plambda}; the frequencies obey $f^4 - (\beta^2 - \gamma^2 + 2\omega^2) f^2 + \omega^4 = 0$, thus
\be \label{evs3}
f_k=\pm\frac{1}{\sqrt{2}} \sqrt{\beta^2 - \gamma^{2} +2 \omega^{2}\pm \sqrt{(\beta^2 - \gamma^2)(\beta^2 - \gamma^{2} +4\omega^{2})}} ,
\ee
for $ k = 1,2,3,4$; in the unbroken phase these are the real frequencies $\pm f_\pm$ of \rr{splitting}.
In the whispering-gallery protocol the gain-loss
parameter $\gamma$ is kept fixed and the ``coupling'' parameter---here the odd friction $\beta$---is varied. The three regimes of Fig. \ref{fig2} are then traversed, and the variation of the real and imaginary parts of $f$ with $\beta$ is displayed in Fig. \ref{fig: freqsbeta}: for $\beta > \gamma$ the four frequencies are real (unbroken $\mathscr{PT}$, Rabi oscillations); for $\sqrt{\gamma^2 - 4\omega^2}<\beta<\gamma$ they form complex quadruplets; for $\beta < \sqrt{\gamma^2-4\omega^2}$ (which requires $\gamma > 2\omega$) they are purely imaginary. Note that if $\gamma \leq 2\omega$ the region with four imaginary $f$ disappears.

\section{\label{sec: lindbladapp}Moment equations of the quantum loss-gain pair}
From the master equation \rr{lindblad}, $\frac{d}{dt}\langle \hat O\rangle = \frac{i}{\hbar}\langle [\hat H, \hat O]\rangle + \gamma \langle \mathcal{D}^\dagger[\hat a]\hat O\rangle + \gamma\langle \mathcal{D}^\dagger[\hat b^\dagger]\hat O\rangle$ with $\mathcal{D}^\dagger[\hat L]\hat O = \hat L^\dagger \hat O \hat L - \frac12\{\hat L^\dagger \hat L, \hat O\}$. For $\hat O \in \{\hat a^\dagger\hat a,\; \hat b^\dagger \hat b,\; \hat a^\dagger \hat b + \hat b^\dagger \hat a\}$ the commutators with $\hat H_c = i\hbar\frac{\beta}{2}(\hat a^\dagger \hat b - \hat a\hat b^\dagger)$ close on the same set, the loss channel contributes $-\gamma n_a$ and $-\frac{\gamma}{2}s$, and the gain channel contributes $+\gamma(n_b+1)$ and $+\frac{\gamma}{2}s$; the two $\pm\frac{\gamma}{2}s$ contributions cancel in $\dot s$, producing \rr{moments}. The homogeneous matrix
\be
M = \begin{pmatrix} -\gamma & 0 & \beta/2 \\ 0 & \gamma & -\beta/2 \\ -\beta & \beta & 0\end{pmatrix}
\ee
has eigenvalues $\{0, \pm\sqrt{\gamma^2 - \beta^2}\}$. The left null vector $\mathbf{w} = (1, 1, -\gamma/\beta)$ satisfies $\mathbf{w}M = 0$, so that $\frac{d}{dt}\left[ n_a + n_b - \frac{\gamma}{\beta} s \right] = \mathbf{w}\cdot(0,\gamma,0)^T = \gamma$ exactly, for all $\beta, \gamma$. In the unbroken phase ($\beta>\gamma$) the remaining eigenvalues are imaginary, $s$ and $n_a - n_b$ remain bounded and oscillate at frequency $\sqrt{\beta^2-\gamma^2}$, and the total occupation therefore grows linearly, $\langle N(t)\rangle = \gamma t + \textrm{(bounded oscillations)}$; in the broken phase the positive eigenvalue $\sqrt{\gamma^2-\beta^2}$ dominates and the growth is exponential, Eq. \rr{heating}.

\bibliographystyle{elsarticle-num}
\bibliography{biblio}

@article{Avron1998,
	author = {Avron, J.E.},
	journal = {{Journal of Statistical Physics}},
	number = {3-4},
	pages = {543--557},
	publisher = {Springer},
	title = {Odd viscosity},
	volume = {92},
	year = {1998}}

@article{Avron1995,
	author = {Avron, J.E. and Seiler, R. and Zograf, P.G.},
	journal = {{Physical Review Letters}},
	number = {4},
	pages = {697},
	publisher = {APS},
	title = {{Viscosity of quantum Hall fluids}},
	volume = {75},
	year = {1995}}

@article{Lapa2014,
	author = {Lapa, M.F. and Hughes, T.L.},
	journal = {{Physical Review E}},
	number = {4},
	pages = {043019},
	publisher = {APS},
	title = {{Swimming at low Reynolds number in fluids with odd, or Hall, viscosity}},
	volume = {89},
	year = {2014}}

@article{Ganeshan2017,
	author = {Ganeshan, S. and Abanov, A.G.},
	journal = {Physical Review Fluids},
	number = {9},
	pages = {094101},
	publisher = {APS},
	title = {Odd viscosity in two-dimensional incompressible fluids},
	volume = {2},
	year = {2017}}

@article{Srivastava2024,
	author = {Srivastava, S. and Monteiro, G.M. and Ganeshan, S.},
	journal = {arXiv preprint arXiv:2406.04570},
	title = {Modulation instability in dispersive parity-broken systems},
	year = {2024}}

@article{Soni2019,
	author = {Soni, V. and Bililign, E.S. and Magkiriadou, S. and Sacanna, S. and Bartolo, D. and Shelley, M.J. and Irvine, W.T.M.},
	journal = {Nature Physics},
	number = {11},
	pages = {1188--1194},
	publisher = {Nature Publishing Group},
	title = {The odd free surface flows of a colloidal chiral fluid},
	volume = {15},
	year = {2019}}

@article{Bateman1931,
	author = {Bateman, H.},
	journal = {{Physical Review}},
	number = {4},
	pages = {815},
	publisher = {APS},
	title = {On dissipative systems and related variational principles},
	volume = {38},
	year = {1931}}

@book{Morse1953,
	address = {N.Y.},
	author = {Morse, P.M. and Feshbach, H.},
	publisher = {McGraw Hill Book Company Inc.},
	series = {International Series in Pure and Applied Physics},
	title = {Methods of Theoretical Physics},
	year = {1953}}

@book{Bender2019,
	author = {Bender, C. M.},
	publisher = {World Scientific},
	title = {{PT symmetry: In quantum and classical physics}},
	year = {2019}}

@article{Bender2013,
	author = {Bender, C.M. and Gianfreda, M. and {\"O}zdemir, {\c{S}}.K. and Peng, B. and Yang, L.},
	journal = {{Physical Review A}},
	number = {6},
	pages = {062111},
	publisher = {APS},
	title = {Twofold transition in {PT}-symmetric coupled oscillators},
	volume = {88},
	year = {2013}}

@article{Peng2014,
	author = {Peng, B. and {\"O}zdemir, {\c{S}}. K. and Lei, F. and Monifi, F. and Gianfreda, M. and Long, G. L. and Fan, S. and Nori, F. and Bender, C.M. and Yang, L.},
	journal = {Nature Physics},
	number = {5},
	pages = {394--398},
	publisher = {Nature Publishing Group UK London},
	title = {Parity--time-symmetric whispering-gallery microcavities},
	volume = {10},
	year = {2014}}

@article{Caprini2025,
	author = {Caprini, L. and Marini Bettolo Marconi, U.},
	journal = {New Journal of Physics},
	number = {5},
	pages = {054401},
	publisher = {IOP Publishing},
	title = {Odd active solids: vortices, velocity oscillations and dissipation-free modes},
	volume = {27},
	year = {2025}}

@article{Dekker1981,
	author = {Dekker, H.},
	journal = {Physics Reports},
	number = {1},
	pages = {1--110},
	publisher = {Elsevier},
	title = {Classical and quantum mechanics of the damped harmonic oscillator},
	volume = {80},
	year = {1981}}

@article{kirkinis2024,
	author = {Kirkinis, E. and Olvera de la Cruz, M.},
	journal = {{Journal of Fluid Mechanics}},
	pages = {A13},
	publisher = {Cambridge University Press},
	title = {Evanescent and inertial-like waves in rigidly rotating odd viscous liquids},
	volume = {996},
	year = {2024}}

@article{Kirkinis2023taylor,
	author = {Kirkinis, E. and Olvera de la Cruz, M.},
	journal = {{Journal of Fluid Mechanics}},
	pages = {A30},
	publisher = {Cambridge University Press},
	title = {Taylor columns and inertial-like waves in a three-dimensional odd viscous liquid},
	volume = {973},
	year = {2023}}

@book{Chadwick1976,
	author = {Chadwick, P.},
	mrclass = {73.00 (76.00)},
	mrnumber = {MR0388914 (52 \#9746)},
	note = {Concise theory and problems},
	pages = {174},
	publisher = {Halsted Press [John Wiley \& Sons], New York (Now in Dover)},
	title = {Continuum mechanics},
	year = {1976}}

@inbook{Serrin1959,
	address = {Berlin, Heidelberg},
	author = {Serrin, James},
	booktitle = {Fluid Dynamics I / Str{\"o}mungsmechanik I},
	editor = {Truesdell, C.},
	isbn = {978-3-642-45914-6},
	pages = {125--263 of Fluid Dynamics I / Str{\"o}mungsmechanik I, Handbuch der Physik},
	publisher = {Springer},
	title = {Mathematical Principles of Classical Fluid Mechanics},
	year = {1959},
	}

@article{Muller1995,
	author = {M{\"u}ller, P.},
	journal = {{Reviews of Geophysics}},
	number = {1},
	pages = {67--97},
	publisher = {Wiley Online Library},
	title = {Ertel's potential vorticity theorem in physical oceanography},
	volume = {33},
	year = {1995}}

@book{Acheson1990,
	address = {New York},
	author = {Acheson, D. J.},
	isbn = {0-19-859660-X; 0-19-859679-0},
	mrclass = {76-01},
	mrnumber = {MR1069557 (93k:76001)},
	mrreviewer = {Andrei Iacob},
	pages = {x+397},
	publisher = {The Clarendon Press Oxford University Press},
	series = {Oxford Applied Mathematics and Computing Science Series},
	title = {Elementary fluid dynamics},
	year = {1990}}

@book{Pedlosky1987,
	author = {Pedlosky, J.},
	edition = {2nd},
	publisher = {Springer-Verlag New York Inc.},
	title = {Geophysical Fluid Dynamics},
	year = {1987}}

@book{Landau1987,
	author = {Landau, L. D. and Lifshitz, E. M.},
	mrclass = {81.0X},
	mrnumber = {MR0093319 (19,1230k)},
	pages = {xii+515},
	publisher = {Pergamon Press Ltd., London-Paris},
	title = {Fluid Mechanics. {C}ourse of {T}heoretical {P}hysics, {V}ol. 6},
	year = {1987}}

@article{kirkinis2019b,
	author = {Kirkinis, E. and Andreev, A.V.},
	journal = {Journal of Fluid Mechanics},
	pages = {169--189},
	publisher = {Cambridge University Press},
	title = {{Odd viscosity-induced stabilization of viscous thin liquid films}},
	volume = {878},
	year = {2019}}

@article{Banerjee2017,
	author = {Banerjee, D. and Souslov, A. and Abanov, A.G. and Vitelli, V.},
	journal = {Nature Communications},
	pages = {1573},
	title = {Odd viscosity in chiral active fluids},
	volume = {8},
	year = {2017}}

@article{Fruchart2023,
	author = {Fruchart, M. and Scheibner, C. and Vitelli, V.},
	journal = {Annual Review of Condensed Matter Physics},
	pages = {471--510},
	title = {Odd viscosity and odd elasticity},
	volume = {14},
	year = {2023}}

@article{Hosaka2021,
	author = {Hosaka, Y. and Komura, S. and Andelman, D.},
	journal = {{Physical Review E}},
	number = {4},
	pages = {042610},
	title = {Nonreciprocal response of a two-dimensional fluid with odd viscosity},
	volume = {103},
	year = {2021}}

@article{Khain2022,
	author = {Khain, T. and Scheibner, C. and Fruchart, M. and Vitelli, V.},
	journal = {{Journal of Fluid Mechanics}},
	pages = {A23},
	title = {Stokes flows in three-dimensional fluids with odd and parity-violating viscosities},
	volume = {934},
	year = {2022}}

@article{Abanov2018,
	author = {Abanov, A. and Can, T. and Ganeshan, S.},
	journal = {SciPost Physics},
	pages = {010},
	title = {Odd surface waves in two-dimensional incompressible fluids},
	volume = {5},
	year = {2018}}

@article{Salmon1988,
	author = {Salmon, R.},
	journal = {Annual Review of Fluid Mechanics},
	pages = {225--256},
	title = {Hamiltonian fluid mechanics},
	volume = {20},
	year = {1988}}

@article{Nash2015,
	author = {Nash, L.M. and Kleckner, D. and Read, A. and Vitelli, V. and Turner, A.M. and Irvine, W.T.M.},
	journal = {Proceedings of the National Academy of Sciences},
	number = {47},
	pages = {14495--14500},
	title = {Topological mechanics of gyroscopic metamaterials},
	volume = {112},
	year = {2015}}

@article{Wiersig2014,
	author = {Wiersig, J.},
	journal = {{Physical Review Letters}},
	number = {20},
	pages = {203901},
	title = {Enhancing the sensitivity of frequency and energy splitting detection by using exceptional points},
	volume = {112},
	year = {2014}}

@article{Chen2017,
	author = {Chen, W. and {\"O}zdemir, {\c{S}}.K. and Zhao, G. and Wiersig, J. and Yang, L.},
	journal = {Nature},
	pages = {192--196},
	title = {Exceptional points enhance sensing in an optical microcavity},
	volume = {548},
	year = {2017}}

@article{ElGanainy2018,
	author = {El-Ganainy, R. and Makris, K.G. and Khajavikhan, M. and Musslimani, Z.H. and Rotter, S. and Christodoulides, D.N.},
	journal = {Nature Physics},
	pages = {11--19},
	title = {Non-{Hermitian} physics and {PT} symmetry},
	volume = {14},
	year = {2018}}

@article{Read2009,
	author = {Read, N.},
	journal = {{Physical Review B}},
	number = {4},
	pages = {045308},
	title = {Non-{Abelian} adiabatic statistics and {Hall} viscosity in quantum Hall states and $p_x+ip_y$ paired superfluids},
	volume = {79},
	year = {2009}}

@article{ReadRezayi2011,
	author = {Read, N. and Rezayi, E.H.},
	journal = {{Physical Review B}},
	number = {8},
	pages = {085316},
	title = {Hall viscosity, orbital spin, and geometry: paired superfluids and quantum {Hall} systems},
	volume = {84},
	year = {2011}}

@article{HoyosSon2012,
	author = {Hoyos, C. and Son, D.T.},
	journal = {{Physical Review Letters}},
	number = {6},
	pages = {066805},
	title = {Hall viscosity and electromagnetic response},
	volume = {108},
	year = {2012}}

@article{Hoyos2014,
	author = {Hoyos, C.},
	journal = {International Journal of Modern Physics B},
	number = {15},
	pages = {1430007},
	title = {Hall viscosity, topological states and effective theories},
	volume = {28},
	year = {2014}}

@article{Bradlyn2012,
	author = {Bradlyn, B. and Goldstein, M. and Read, N.},
	journal = {{Physical Review B}},
	number = {24},
	pages = {245309},
	title = {{Kubo formulas for viscosity: Hall viscosity, Ward identities, and the relation with conductivity}},
	volume = {86},
	year = {2012}}

@article{Berdyugin2019,
	author = {Berdyugin, A.I. and Xu, S.G. and Pellegrino, F.M.D. and Krishna Kumar, R. and Principi, A. and Torre, I. and Ben Shalom, M. and Taniguchi, T. and Watanabe, K. and Grigorieva, I.V. and Polini, M. and Geim, A.K. and Bandurin, D.A.},
	journal = {Science},
	number = {6436},
	pages = {162--165},
	title = {Measuring {Hall} viscosity of graphene's electron fluid},
	volume = {364},
	year = {2019}}

@article{Pellegrino2017,
	author = {Pellegrino, F.M.D. and Torre, I. and Polini, M.},
	journal = {{Physical Review B}},
	number = {19},
	pages = {195401},
	title = {Nonlocal transport and the {Hall} viscosity of two-dimensional hydrodynamic electron liquids},
	volume = {96},
	year = {2017}}

@article{Lucas2018,
	author = {Lucas, A. and Fong, K.C.},
	journal = {Journal of Physics: Condensed Matter},
	number = {5},
	pages = {053001},
	title = {Hydrodynamics of electron fluids in graphene},
	volume = {30},
	year = {2018}}

@article{Fock1928,
	author = {Fock, V.},
	journal = {Zeitschrift f{\"u}r Physik},
	pages = {446--448},
	title = {Bemerkung zur {Q}uantelung des harmonischen {O}szillators im {M}agnetfeld},
	volume = {47},
	year = {1928}}

@article{Darwin1930,
	author = {Darwin, C.G.},
	journal = {Mathematical Proceedings of the Cambridge Philosophical Society},
	number = {1},
	pages = {86--90},
	title = {The diamagnetism of the free electron},
	volume = {27},
	year = {1930}}

@article{Tarucha1996,
	author = {Tarucha, S. and Austing, D.G. and Honda, T. and van der Hage, R.J. and Kouwenhoven, L.P.},
	journal = {{Physical Review Letters}},
	number = {17},
	pages = {3613},
	title = {Shell filling and spin effects in a few electron quantum dot},
	volume = {77},
	year = {1996}}

@article{Dalibard2011,
	author = {Dalibard, J. and Gerbier, F. and Juzeli{\=u}nas, G. and {\"O}hberg, P.},
	journal = {Reviews of Modern Physics},
	number = {4},
	pages = {1523},
	title = {Colloquium: Artificial gauge potentials for neutral atoms},
	volume = {83},
	year = {2011}}

@article{Schine2016,
	author = {Schine, N. and Ryou, A. and Gromov, A. and Sommer, A. and Simon, J.},
	journal = {Nature},
	pages = {671--675},
	title = {Synthetic {Landau} levels for photons},
	volume = {534},
	year = {2016}}

@article{Scheel2018,
	author = {Scheel, S. and Szameit, A.},
	journal = {{EPL (Europhysics Letters)}},
	number = {3},
	pages = {34001},
	title = {{PT}-symmetric photonic quantum systems with gain and loss do not exist},
	volume = {122},
	year = {2018}}

@article{Minganti2019,
	author = {Minganti, F. and Miranowicz, A. and Chhajlany, R.W. and Nori, F.},
	journal = {{Physical Review A}},
	number = {6},
	pages = {062131},
	title = {Quantum exceptional points of non-{H}ermitian {H}amiltonians and {L}iouvillians: The effects of quantum jumps},
	volume = {100},
	year = {2019}}

@article{Hargus2021,
	author = {Hargus, C. and Epstein, J.M. and Mandadapu, K.K.},
	journal = {{Physical Review Letters}},
	number = {17},
	pages = {178001},
	title = {Odd diffusivity of chiral random motion},
	volume = {127},
	year = {2021}}

@article{Han2021,
	author = {Han, M. and Fruchart, M. and Scheibner, C. and Vaikuntanathan, S. and de Pablo, J.J. and Vitelli, V.},
	journal = {Nature Physics},
	pages = {1260--1269},
	title = {Fluctuating hydrodynamics of chiral active fluids},
	volume = {17},
	year = {2021}}

@article{Aggarwal2023,
  title={Thermocapillary migrating odd viscous droplets},
  author={Aggarwal, A. and Kirkinis, E. and Olvera de la Cruz, M.},
  journal={{Physical Review Letters}},
  volume={131},
  number={19},
  pages={198201},
  year={2023},
  publisher={APS}
}

@article{Guo2009,
  title={{Observation of PT-symmetry breaking in complex optical potentials}},
  author={Guo, A. and Salamo, G. J. and Duchesne, D. and Morandotti, R. and Volatier-Ravat, M. and Aimez, V. and Siviloglou, G.A. and Christodoulides, D.N.},
  journal={{Physical Review Letters}},
  volume={103},
  number={9},
  pages={093902},
  year={2009},
  publisher={APS}
}

@article{Lin2011,
  title={{Unidirectional invisibility induced by PT-symmetric periodic structures}},
  author={Lin, Z. and Ramezani, H. and Eichelkraut, T. and Kottos, T. and Cao, H. and Christodoulides, D.N.},
  journal={{Physical Review Letters}},
  volume={106},
  number={21},
  pages={213901},
  year={2011},
  publisher={APS}
}

@book{Landau1981,
  title={{Course of theoretical physics. {V}ol. 10: Physical Kinetics}},
  author={Lifshitz, E. M. and Pitaevskii, L. P.},
  year={1981},
  publisher={Pergamon Press}
}

@article{Millikan1929,
  title={On the steady motion of viscous, incompressible fluids; with particular reference to a variation principle},
  author={Millikan, C.B.},
  journal={Philosophical Magazine},
  volume={7},
  number={44},
  pages={641--662},
  year={1929}
}

@article{Finlayson1972,
  title={Existence of variational principles for the {N}avier-{S}tokes equation},
  author={Finlayson, B.A.},
  journal={Physics of Fluids},
  volume={15},
  number={6},
  pages={963--967},
  year={1972}
}

@book{Olver1993,
  title={Applications of {L}ie Groups to Differential Equations},
  author={Olver, P.J.},
  series={Graduate Texts in Mathematics},
  volume={107},
  edition={2nd},
  publisher={Springer},
  address={New York},
  year={1993}
}

@article{Morandi1990,
  title={The inverse problem in the calculus of variations and the geometry of the tangent bundle},
  author={Morandi, G. and Ferrario, C. and Lo Vecchio, G. and Marmo, G. and Rubano, C.},
  journal={Physics Reports},
  volume={188},
  number={3-4},
  pages={147--284},
  year={1990}
}

@article{SeligerWhitham1968,
  title={Variational principles in continuum mechanics},
  author={Seliger, R.L. and Whitham, G.B.},
  journal={Proceedings of the Royal Society of London A},
  volume={305},
  pages={1--25},
  year={1968}
}

@article{AbanovMonteiro2019,
  title={Free-surface variational principle for an incompressible fluid with odd viscosity},
  author={Abanov, A.G. and Monteiro, G.M.},
  journal={{Physical Review Letters}},
  volume={122},
  number={15},
  pages={154501},
  year={2019}
}

@misc{Claude2026,
	author = {{Anthropic}},
	howpublished = {\url{https://claude.ai}},
	note = {Version: Claude Fable 5; used June--July 2026},
	title = {Claude [Large language model]},
	year = {2026}}

@article{Haynes1990,
  title={On the conservation and impermeability theorems for potential vorticity},
  author={Haynes, P.H. and McIntyre, M.E.},
  journal={{Journal of Atmospheric Sciences}},
  volume={47},
  number={16},
  pages={2021--2031},
  year={1990}
}

@article{Haynes1987,
  title={On the evolution of vorticity and potential vorticity in the presence of diabatic heating and frictional or other forces},
  author={Haynes, P.H. and McIntyre, M.E.},
  journal={{Journal of the Atmospheric Sciences}},
  volume={44},
  number={5},
  pages={828--841},
  year={1987},
  publisher={American Meteorological Society}
}

@article{Kirkinis2025wave,
  title={Wave-crests around obstacles in odd viscous liquids},
  author={Kirkinis, E. and Levchenko, A.},
  journal={arXiv preprint arXiv:2511.07751},
  year={2025}
}

@article{Kirkinis2023halos,
    author = {Kirkinis, E. and Olvera de la Cruz, M.},
    title = "{Taylor halos and Taylor spears in odd viscous liquids}",
    journal = {Physics of Fluids},
    volume = {35},
    number = {10},
    pages = {101702},
    year = {2023},
    month = {10},
    issn = {1070-6631},
}

\end{document}